\documentclass[aps,preprint,nofootinbib]{revtex4}%
\usepackage{amssymb}
\usepackage{amsfonts}
\usepackage{amsmath}
\usepackage{graphicx}%
\providecommand{\U}[1]{\protect\rule{.1in}{.1in}}

\usepackage[bbgreekl]{mathbbol}
\usepackage{array}

\begin{document}
\preprint{ }
\title{Signals of Detailed Balance Violation in Nonequilibrium Stationary States:
Subtle, Manifest, and Extraordinary}
\author{R. K. P. Zia}

\email[E-mail: ]{rkpzia@vt.edu}

\affiliation{
Center for Soft Matter and Biological Physics, Department of Physics,
Virginia Polytechnic Institute \& State University, Blacksburg, VA 24061, USA
}

\date{February 23, 2024}

\begin{abstract}
The evolution of physical systems are often modeled by simple Markovian
processes. When settled into stationary states, the probability distributions
of such systems are time independent, by definition. However, they do not
necessarily fall within the framework of equilibrium statistical mechanics.
Instead, they may be \textit{non-equilibrium} steady states (NESS). One
distinguished feature of NESS is the presence of time reversal asymmetry (TRA)
and persistent probability current loops. These loops lead naturally to the
notion of probability angular momenta, which play a role on the same footing
as the noise covariance in stochastic processes.
Illustrating with simulations of simple models and physical data, we present
ways to detect these signals of TRA, from the subtle to the prominent.

\end{abstract}
\maketitle

\section{Introduction}

\label{I}To formulate a theory of equilibrium statistical mechanics, the
principle of detailed balance (DB) played a foundational role for both
Boltzmann and Maxwell. Based on the laws of classical physics, it embodies the
notion of time reversal invariance. The key theme is that, once it
\textquotedblleft settled down into a stationary state\footnote{Note that the
definition of a \textit{stationary} state involves time \textit{translational
invariance}. Time \textit{reversal} is a different symmetry.}%
,\textquotedblright\ the evolution of an isolated system is
(statistically)\ the same whether time is reversed or not. A favorite way to
put this notion in layman's terms is the following: If you are shown a movie
of a system in thermal equilibrium, you cannot tell if the movie is run
forwards or backwards. A more sophisticated way to describe this state is that
its entropy has reached its maximum and cannot increase further.

If a system is not in isolation, but in contact with multiple reservoirs which
are not in equilibrium, then the dynamics of these reservoirs may lead to a
flow of energy (or matter, or information, or ...) \textit{through }our
system. Good examples include all forms of life on Earth as well as our entire
ecosystem. All life forms take in nutrients and discard waste, while our
ecosystem is sustained by a balance between incoming solar radiation and 
outgoing IR rays. When the time-scales of these reservoirs are long enough 
(and spatial scales large enough), then it is conceivable for our 
(comparatively small) system to settle into a 
\textquotedblleft quasi-stationary state\textquotedblright\ 
(i.e., stationary over much shorter time-scales\footnote{All natural 
systems are bound by finite times, which
motivates the term \textquotedblleft quasi-stationary.\textquotedblright\ In
reality, we need systems with a clear separation of time scales so that they
can be observed over periods which are long compared to the \textquotedblleft
microscopic\textquotedblright\ time scales that characterize evolution towards
the stationary state. The present overall state of the sun might be a good
example of quasi-stationarity, as it appears to be \textquotedblleft the
same\textquotedblright\ on the scale of billion-years. By contrast, the years
it took to reach the main sequence were far fewer.}). 
Clearly, it is
prohibitively challenging to describe such a system as well as its
\textquotedblleft environment\textquotedblright\ -- the large, slowly varying
reservoirs. An approximate description is to focus on our system
\textit{alone} and model its evolution with a stochastic dynamics that is
\textit{not} time-reversal invariant, i.e., with DB violating 
rules\footnote{ Indeed, may stochastic models of natural phenomena are of 
this type, e.g., predator-prey, birds flocking, epidemic spreading, 
vehicular traffic, the stock market, etc.}. When such
a system settles down, it will be in a \textit{non-equilibrium steady state}
(NESS), while a movie of it run forwards should be distinguishable from one
run backwards. What are the signals of such time-reversal
asymmetry (TRA)? We will present both generally applicable and comprehensible
results, as well as specific examples which range from the readily solvable
to the most challenging.

To discuss stochastic processes in general is beyond the scope of this
article. Instead, we limit ourselves here to noisy systems that can be modeled
by Markovian dynamics. Starting with the conceptually easy Langevin approach
for a single particle, as well as the equivalent Fokker-Planck description, we
will move on to arbitrary systems with finite and discrete configurations
($\mathcal{C}_{i};~i=1,...,N$) evolving stochastically in discrete time steps
($\tau$) with time independent rates $R\left(  i|j\right)  $. There, the focus
will be on the distribution $P\left( i,\tau\right)  $ -- the probability to
find our system in $\mathcal{C}_{i}$ at time $\tau$ -- given an initial
condition $P\left(  k,0\right)  $. By limiting ourselves this way, the full
evolution can be specified by a Master equation for $P$, based on a given set
$R$'s. The next Section will be devoted\footnote{With pedagogy in mind, this 
section is written mainly for students unfamiliar with the various approaches.} 
to these different set ups. As a
linear equation for a finite number of variables, the Master equation is, in
principle, solvable. The remainder of this article consists of expanding on
the notion of DB, the existence of non-trivial probability currents, and TRA
in NESS. Generalities will be presented in in Section III, with the main focus
on probability angular momenta. Its relationship with ordinary angular momenta 
and, remarkably, the covariance of the noise, will be highlighted in 
Subsection \ref{formal}. 
Though most of this article will focus on probability angular momenta, 
a short digression will be introduced (Subsection \ref{sec:1D}) to discuss 
\textquotedblleft one dimensional\textquotedblright\ systems,  
which do not ordinarily support angular momenta.
In Section \ref{Examples}, we present a number of
illustrations, from simple simulation to physical data. They show a range of
systems which display signals of DB violation and TRA, from the
\textquotedblleft subtle\textquotedblright\ to the \textquotedblleft
manifest.\textquotedblright\ Beyond, we emphasize that DB violation can also
lead to NESS with astonishingly unexpected phenomena, such as all living
organisms. This level of \textquotedblleft extraordinary\textquotedblright%
\ behavior will be illustrated in a simple example -- one recently 
studied\cite{DZ18,LDZ21} in the context of driven diffusive systems. 
Several appendices contain much of the mathematical details of our
discussions. We end with a brief summary and outlook. 

\section{Equations for stochastic processes and stationary distributions
$P^{\ast}$}

\label{II}There are several equivalent ways to describe a stochastic process.
Conceptually, arguably the simplest is the Langevin approach\cite{Langevin},
where noise is incorporated into equations of evolution for the degrees of
freedom of interest. In the example of a single particle moving in ordinary
space-time $\left(  \vec{x},t\right)  $, the Langevin equation reads\footnote{
For simplicity, we restrict ourselves specifically to additive noise 
in the Ito formulation. Generalizations are clearly possible.}
$m\partial_{t}^{2}\vec{x}=\vec{F}+\vec{\eta}$. Here $\vec{\eta}$ is a random
force, often chosen as Gaussian distributed, with zero mean and constant
variance, uncorrelated from one time to another. The histories, $\vec
{x}\left(  t\right)  $'s, resulting from different realizations of $\vec{\eta
}$ form \textquotedblleft spaghetti plots\textquotedblright\ (as often seen in
hurricane forecasts), which provide a rough sense of $P(\vec{x},t)$, the
probability of finding it at $\vec{x}$ at time $t$. An alternative is the
deterministic Fokker-Planck equation\cite{Fokker} for $P$, while standard
methods link one description to the other\footnote{The derivation from one to
the other can be found in standard references, such as those listed in
https://en.wikipedia.org/wiki/Fokker\%E2\%80\%93Planck\_equation.}. The
simplest version of the former (representing the over-damped, $m\rightarrow0$
limit of $\vec{F}=m\vec{a}$) is\footnote{Vectors in continuous spaces (with a
metric) are denoted in the usual fashion, e.g., $\vec{x}$ for position in
ordinary 3D space. Components will carry a Greek index, e.g., $x_{\alpha}$.
Matrices will be denoted by blackboard bold font, e.g., $\mathbb{M}$. In this
setup, the noise correlation reads $\left\langle \eta_{\alpha}\left(
t\right)  \eta_{\beta}\left(  t^{\prime}\right)  \right\rangle =2D_{\alpha
\beta}\delta\left(  t-t^{\prime}\right)  $ in component form, or 
written compactly as 
$\left\langle \vec{\eta}\vec{\eta}^{T}\right\rangle \propto2\mathbb{D}$.}%
\begin{equation}
\partial_{t}\vec{x}=\vec{\mu}+\vec{\eta};~~\left\langle \vec{\eta}\vec{\eta
}^{T}\right\rangle \propto2\mathbb{D} \label{LE}%
\end{equation}
It is equivalent to drift with diffusion in the latter:
\begin{equation}
\partial_{t}P=-\vec{\nabla}\cdot\left[  \vec{\mu}P-\mathbb{D}\vec{\nabla
}P\right]  \label{FPE}%
\end{equation}
In these approaches, the particle is allowed to take only infinitesimal steps
in both $\vec{x}$ and $t$. Obviously, these equations can be generalized to
many variables (to be denoted by $\xi_{\alpha}$ below) evolving in continuous
$t$.

The Master equation is a generalization, in that all type of
stochastic processes can be modeled, beyond the bounds of infinitesimal steps
in continuous space-time. For example, we may consider the dollar value of a
stock from one trading session to the next, the number of individuals infected
with COVID-19 from week to week, or the state of $L^{d}$ spins of an Ising
model in a $d$-dimensional\ cubic lattice, flipping in a Monte-Carlo simulation.

To keep the presentation here simple and to relate directly to simulation
studies, let us consider systems with \textit{discrete} configurations
$\mathcal{C}_{i}$ evolving in \textit{discrete} time steps (of unit
$\varepsilon$). For short, we will use $i$ and $\tau=0,\varepsilon
,2\varepsilon,...$ (in lieu of continuous $\vec{x},t$). The stochastic
evolution of the distribution $P\left(  i,\tau\right)  $ is completely
specified by an initial $P\left(  k,0\right)  $ and the set of
\textquotedblleft rates\textquotedblright\ $R\left(  i|j\right)  \geq0$ for
$i\neq j$. In general, these $R$'s are time dependent. 
But, to keep our problem manageable, we 
will restrict ourselves to $t$-\textit{independent} ones here. The Master
equation specifies the change in $P$ from one $\tau$ to the next:%
\begin{align}
\Delta_{\tau}P\left(  i,\tau\right)   &  \equiv P\left(  i,\tau+\varepsilon
\right)  -P\left(  i,\tau\right) \label{ME}\\
&  =\sum_{j\neq i}\left[  R\left(  i|j\right)  P\left(  j,\tau\right)
-R\left(  j|i\right)  P\left(  i,\tau\right)  \right]  \label{ME2}%
\end{align}
It is clear that the two terms in Eqn.(\ref{ME2}) represent the
\textquotedblleft flow of probability into\ $\mathcal{C}_{i}$%
\textquotedblright\ and \textquotedblleft out from $\mathcal{C}_{i}%
$,\textquotedblright\ respectively. Obviously, $R\left(  i|j\right)  $ can be
regarded as the\textit{ conditional probability} for finding $i$ given $j$,
while $R\left(  i|j\right)  P\left(  j,\tau\right)  $ is just the
\textit{joint probability} $\mathcal{P}\left(  i,\tau+\varepsilon\cap
j,\tau\right)  $ for finding the system in configuration $\mathcal{C}_{j}$ at
$\tau$ and in $\mathcal{C}_{i}$ at $\tau+\varepsilon$. Note that, in this
formulation, there is no mention of a metric in $\mathcal{C}$ space, and there
is no restriction on how \textquotedblleft close\textquotedblright%
\ $\mathcal{C}_{i}$ and $\mathcal{C}_{j}$ must be. Unlike the Fokker-Planck
equation, the Master equation can describe a system that may \textquotedblleft
jump\textquotedblright\ from any $\mathcal{C}$ to any other one in the system.
By introducing some notion of distance in $\mathcal{C}$ space (e.g., ordinary
space $\vec{x}$) and allowing only \textquotedblleft infinitesimally close
moves\textquotedblright\ in infinitesimal time steps,\ we can recover a
Fokker-Planck equation from the Master equation by taking appropriate
continuum limits.

Next, we turn to the problem of \textquotedblleft solving\textquotedblright%
\ these equations. Being a stochastic differential equation, the solution to
the Langevin equation, $\vec{x}\left(  t\right)  $, will be different for each
realization of the noise $\vec{\eta}\left(  t\right)  $. In principle,
statistical analysis, using a given distribution for the noise, needs to be
performed to obtain averages and correlations like $\left\langle \vec
{x}\left(  t\right)  \right\rangle $ and $\left\langle \vec{x}\left(
t\right)  \vec{x}\left(  t^{\prime}\right)  \right\rangle $. The Fokker-Planck
equation for $P$ is deterministic and \textit{linear} (like the
Schr\"{o}dinger equation for $\Psi(\vec{x},t)$) and finding $P(\vec{x},t)$ in
general is as difficult as solving for $\Psi(\vec{x},t)$ in quantum mechanics. 
Formally, the
approach in non-relativistic quantum mechanics can be followed, by writing the
right hand side of Eqn.(\ref{FPE}) as an operator -- the Liouvillian
$\mathfrak{L}$ -- on $P$. Then the solution $P\left(  t\right)  $, given some
initial distribution, $P\left(  0\right)  $, is just $\exp\left[
\mathfrak{L}t\right]  P\left(  0\right)  $. The same strategy can be applied
to the Master equation (\ref{ME}). Regarding $P\left(  i,\tau\right)  $ as a
ket $\left\vert P_{\tau}\right\rangle $, we can write the equation in a
compact form\footnote{\textquotedblleft Vectors\textquotedblright\ in
$\mathcal{C}$ space, which may not have a metric, will be denoted by bras and
kets, with $i^{th}$ component of $\left\vert P\right\rangle $ being $P\left(
i\right)  $. Matrices acting on them will be denoted by Fraktur type, e.g.,
$\mathfrak{L}$.}:%
\[
\Delta_{\tau}\left\vert P\right\rangle =\mathfrak{L}\left\vert P\right\rangle
\implies\left\vert P_{\tau+\varepsilon}\right\rangle =\left(  \mathfrak{I}%
+\mathfrak{L}\right)  \left\vert P_{\tau}\right\rangle
\]
where $\mathfrak{I}$\ is the identity matrix. Here, the off-diagonal elements
of $\mathfrak{L}$ are the $R$'s, while the $i^{th}$ diagonal element is the
sum $[ -\sum_{j\neq i}R\left(  j|i\right)  ] $. Given some initial
distribution $\left\vert P_{ini}\right\rangle $, the full solution is just%
\[
\left\vert P_{\tau}\right\rangle =\left(  \mathfrak{I}+\mathfrak{L}\right)
^{\tau}\left\vert P_{ini}\right\rangle
\]
Of course, this expression is somewhat formal and a practical solution will
require finding all the eigenvalues and eigenvectors (both right- and
left-eigenvectors, or kets and bras) of $\mathfrak{L}$ :%
\[
\mathfrak{L}\left\vert u_{A}\right\rangle =\lambda_{A}\left\vert
u_{A}\right\rangle ;~~\left\langle w_{A}\right\vert \mathfrak{L}=\lambda
_{A}\left\langle w_{A}\right\vert
\]
so that $\left\vert P_{\tau}\right\rangle =\sum_{A}\left\vert u_{A}%
\right\rangle \left(  1+\lambda_{A}\right)  ^{\tau}\left\langle w_{A}%
\right\vert \left.  P_{ini}\right\rangle $.

In general, such a task would be prohibitively difficult. However, as
$\mathfrak{L}$ describes a stochastic process for probabilities, it has
special properties: First, since probability is conserved, $\Delta_{\tau}%
\sum_{i}\left\vert P\right\rangle \equiv0$, so that $\left\langle
w_{0}\right\vert =\left(  1,1,1,...,1\right)  $ must be a bra with
$\lambda_{0}\equiv0$. Associated with this is a ket $\left\vert u_{0}%
\right\rangle $, which is readily recognized as a \textit{steady state}, as it
does not change from one $\tau$ to the next. Denoting $\left\vert
u_{0}\right\rangle $ by $\left\vert P^{\ast}\right\rangle $, it is naturally
normalized by $1=$ $\left\langle w_{0}\right\vert \left.  P^{\ast
}\right\rangle $. The issue of other zero eigenvalues and uniqueness of the
pair, $\left\langle w_{0}\right\vert $ and $\left\vert u_{0}\right\rangle $,
is not trivial and will not be addressed in this short article. We will
restrict ourselves to systems in which $\left\vert P^{\ast}\right\rangle $ is
unique. Next, thanks to Perron-Frobenius \cite{PF-wiki}, all $\lambda$'s
besides $\lambda_{0}$ have real parts in $\left[  -1,0\right)  $, so that
$\left\vert P\right\rangle $ will reach $\left\vert P^{\ast}\right\rangle $ at
large times. Much of the rest of this article will focus on this steady state
$\left\vert P^{\ast}\right\rangle $ and its properties.

For $\Delta_{\tau}\left\vert P^{\ast}\right\rangle =0$, it is sufficient (but
not necessary) if each term within $\left[  ...\right]  $ of Eqn.(\ref{ME2})
vanishes:%
\begin{equation}
R\left(  i|j\right)  P^{\ast}\left(  j\right)  =R\left(  j|i\right)  P^{\ast
}\left(  i\right)  \label{DB}%
\end{equation}
a condition often referred to as detailed balance (DB). Note that, in terms of
$\mathcal{P}^{\ast}$, the joint probability in the steady state, this
condition takes the form
\begin{equation}
\mathcal{P}^{\ast}\left(  i,\tau+\varepsilon\cap j,\tau\right)  =\mathcal{P}%
^{\ast}\left(  j,\tau+\varepsilon\cap i,\tau\right)  \label{DBjointP*}%
\end{equation}
The DB condition allows us to exploit computer simulations and generate a set
of configurations that approximate the relative probabilities of a system in
thermal equilibrium (e.g., $P^{\ast}\left(  i\right)  \propto\exp\left\{
-\mathcal{H}\left(  \mathcal{C}_{i}\right)  \right\}  $, where $\mathcal{H}$
is an energy functional). In particular, it suffices to employ any set of
$R$'s that obeys (\ref{DB}). However, if we are only given a set of $R$'s
which specifies a Markovian process, then we typically do not have $P^{\ast
}\left(  i\right)  $ \textit{a priori}, and checking if the $R$'s satisfy DB
or not requires more effort. Not being the focus here, we will just list some
references for the interested reader\cite{W11,Komo36,ZS07}. 

The crux of the issue here is 
that individual terms on the right of Eqn.(\ref{ME2}) need not
vanish, only their sum (at each $i$) must do so. Indeed, since probability is
conserved, each term in the sum is recognized as the \textit{net probability
current}\cite{ZS07}, from $\mathcal{C}_{j}$ to $\mathcal{C}_{i}$: $K\left(
i|j,\tau\right)  \equiv R\left(  i|j\right)  P\left(  j,\tau\right)  -R\left(
j|i\right)  P\left(  i,\tau\right)  $. When the system settles in the steady
state, there is no constraint that the \textit{time-independent }currents%
\begin{align}
K^{\ast}\left(  i|j\right)   &  \equiv R\left(  i|j\right)  P^{\ast}\left(
j\right)  -R\left(  j|i\right)  P^{\ast}\left(  i\right) \label{K*}\\
&  =\mathcal{P}^{\ast}\left(  i,\varepsilon\cap j,0\right)  -\mathcal{P}%
^{\ast}\left(  j,\varepsilon\cap i,0\right)
\end{align}
must vanish. Instead, since $\Delta_{\tau}\left\vert P^{\ast}\right\rangle
=0$, these persistent $K^{\ast}$'s sum to zero at each $i$ and so, they must
form \textit{closed loops.} Such states are analogous to those in
magnetostatics. In this electromagnetic analog, we distinguish
time-independent states by referring those with $K^{\ast}\equiv0$\ (\`{a} la
electrostatics) as \textquotedblleft equilibrium steady
states\textquotedblright\ and those with $K^{\ast}\neq0$ as \textquotedblleft
non-equilibrium steady states\textquotedblright\ -- NESS. While the former can
be achieved by being totally isolated (the micro-canonical ensemble) or being
in equilibrium with one or more, much larger, reservoirs (e.g., energy
reservoir and the canonical ensemble or particle reservoir and the grand
ensemble), the latter are \textit{open} systems, typically coupled to multiple
reservoirs in such a way as to allow a \textit{steady flux} of energy (or
matter, or ...) through them. Such throughput prevents our system from ever
reaching thermal equilibrium and sets up non-trivial, persistent $K^{\ast}$'s
in NESSs.

In the Fokker-Planck formulation, these considerations take a slightly
different form. Conserving probability, Eqn.(\ref{FPE}) is just a continuity
equation, e.g., $\partial_{t}\rho=-\vec{\nabla}\cdot\vec{J}$. Thus, the
probability currents\footnote{To be precise, $\vec{J}_{FP}$'s are probability
current \textit{densities}, just as $\vec{J}$ are current densities. Such 
currents are well-known in non-relativistic quantum mechanics 
(https://en.wikipedia.org/wiki/Probability\_current) and quantum field theory.}
here are \textquotedblleft vector fields\textquotedblright: 
$\vec{J}_{FP}=\vec{\mu}P-\mathbb{D}\vec{\nabla}P$,
with zero divergence in a NESS: $\vec{\nabla}\cdot\vec{J}_{FP}^{\ast}=0$.
Analogous to the usual relation $\vec{J}=\rho\vec{v}$, we may
regard
$\vec{J}%
_{FP}/P=\vec{\mu}-D\vec{\nabla}\ln P$ as the \textquotedblleft probability
velocity field\textquotedblright. However, $\vec{J}_{FP}$ is not the same as
$K$, as the former requires\ the notion of metric on the space of
configurations (e.g., ordinary $\vec{x}$ or the more general $\xi_{\alpha}$)
and transitions to be infinitesimally \textquotedblleft
close\textquotedblright\ (for $\vec{v}$ to be well defined). If we discretize
$\vec{x}$ and write an approximate version of Eqn.(\ref{FPE}) in matrix form,
the presence of $\nabla$ and $\nabla^{2}$ implies that its 
$R\left(  i|j\right)$'s would be zero for $i,j$ being further than nearest 
neighbor pairs, say. In this sense, we regard the Master equation as more
general, with no restrictions on $R$. Though $\vec{J}_{FP}$ and $K$ are
different, the DB condition still takes the form $\vec{J}_{FP}^{\ast}=0$, 
and is a property of $\vec{\mu}$ and $\mathbb{D}$. 
Clearly, $\vec{J}_{FP}^{\ast}\neq0$
in a NESS but must be \textquotedblleft a curl,\textquotedblright\ i.e.,
divergence of an anti-symmetric tensor field. Thus, we have the concept of
current loops here, much like in magnetostatics.

To summarize, systems evolving with DB respecting $R$'s will settled into
equilibrium stationary states with $K^{\ast}\equiv0$ and a readily obtainable
$P^{\ast}$, while systems subjected to DB violating $R$'s will settle into NE
steady states. Generally, given DB violating $R$'s, the distribution $P^{\ast
}$ is unknown (though a formal solution exists \cite{Hill66,ZS07}), while some
non-trivial, persistent current loops will be present. Exploring the
consequences of these $K^{\ast}$'s were initiated sometime ago\cite{ZS07}. The
remainder of this article will be devoted to more recent efforts: detecting
various signals of DB violation in NESS, ranging from quite subtle ones and
clearly self-evident ones to those displaying extraordinary, unexpected phenomena.

\section{Detailed balance violation and time reversal asymmetry in NESS}

\label{III}In a stochastic process, $P\left(  i,\mathcal{\tau}\right)  $ may
seem abstract and unmanageable. Instead, we may focus on expectation values of
observables. Denoting the observables as $\mathcal{O}\left(  i\right)  $ --
functionals of the configurations $\mathcal{C}_{i}$, the expectations at time
$\tau$ are%
\[
\left\langle \mathcal{O}\right\rangle _{\mathcal{\tau}}\equiv\sum
_{i}\mathcal{O}\left(  i\right)  P\left(  i,\mathcal{\tau}\right)
\]
By definition, a system in steady state is invariant under \textit{time}
\textit{translation}, i.e., $\left\langle \mathcal{O}\right\rangle
_{\mathcal{\tau}}=\left\langle \mathcal{O}\right\rangle _{\mathcal{\tau
}+\upsilon}$ for any $\upsilon$ and are denoted by
\[
\left\langle \mathcal{O}\right\rangle ^{\ast}\equiv\sum_{i}\mathcal{O}P^{\ast
}\left(  i\right)
\]
Thanks to this property, simulation studies of steady states often replace
ensemble averages, $\left\langle \cdot\right\rangle ^{\ast}$, by time
averages, $\overline{{\Large \cdot}}$, over a single long run (or very few
runs). In examples of simulation results we present below, $\left\langle
\cdot\right\rangle ^{\ast}$ are obtained from such time averages.

It is clear that, to detect \textit{time reversal asymmetry} in steady states,
we must start with expectations of observables at \textit{different} times,
e.g.,
\begin{equation}
\left\langle \mathcal{O}^{\prime}\mathcal{O}\right\rangle _{\mathcal{\tau
}^{\prime},\mathcal{\tau}}\equiv\sum_{i,i^{\prime}}\mathcal{O}^{\prime}\!\left(
i^{\prime}\right)  \mathcal{O}\left(  i\right)  \mathcal{P}\left(  i^{\prime
},\mathcal{\tau}^{\prime}\cap i,\mathcal{\tau}\right)  \label{2-2}%
\end{equation}
Due to translational invariance, the joint probability, $\mathcal{P}^{\ast
}\left(  i^{\prime},\mathcal{\tau}^{\prime}\cap i,\mathcal{\tau}\right)  $,
depends only on the difference $\upsilon\equiv\mathcal{\tau}^{\prime
}\mathcal{-\tau}$ and we can write
\[
\left\langle \mathcal{O}^{\prime}\mathcal{O}\right\rangle _{\upsilon}^{\ast
}\equiv\sum_{i^{\prime},i}\mathcal{O}^{\prime}\!\left(  i^{\prime}\right)
\mathcal{O}\left(  i\right)  P\left(  i^{\prime},\upsilon|i,0\right)  P^{\ast
}\left(  i\right)
\]
Assuming $\upsilon>0$, the conditional probability, $P\left(  i^{\prime
},\upsilon|i,0\right)  $, is simply the operator $\left(  \mathfrak{I}%
+\mathfrak{L}\right)  ^{\upsilon}$. In particular, if the time difference is
just a single step, we can exploit $P\left(  j,\varepsilon|i,0\right)
=R\left(  j|i\right)  $ and write%
\[
\left\langle \mathcal{O}^{\prime}\mathcal{O}\right\rangle _{\varepsilon}%
^{\ast}\equiv\sum_{j,i}\mathcal{O}^{\prime}\!\left(  j\right)  \mathcal{O}%
\left(  i\right)  R\left(  j|i\right)  P^{\ast}\left(  i\right)
\]

In this setup, time reversal asymmetry (TRA) is present if
\begin{equation}
\left\langle \mathcal{O}^{\prime}\mathcal{O}\right\rangle _{\mathcal{\tau
}^{\prime},\mathcal{\tau}}-\left\langle \mathcal{O}^{\prime}\mathcal{O}%
\right\rangle _{\mathcal{\tau},\mathcal{\tau}^{\prime}}\neq0\label{22TRA}%
\end{equation}
Clearly, a necessary condition is that the two observables are distinct:
$\mathcal{O}\neq\mathcal{O}^{\prime}$. In the steady state, we see that this
quantity is \textit{anti-symmetric} in $\upsilon$:%
\[
\mathcal{A}\left(  \upsilon\right)  \equiv\left\langle \mathcal{O}^{\prime
}\mathcal{O}\right\rangle _{\upsilon}^{\ast}-\left\langle \mathcal{O}^{\prime
}\mathcal{O}\right\rangle _{-\upsilon}^{\ast}%
\]
If we focus on a single time step difference ($\upsilon=\varepsilon$), then
\[
\mathcal{A}\left(  \varepsilon\right)  =\sum_{i,j}\left[  \mathcal{O}^{\prime}\!\left(  j\right)  \mathcal{O}\left(  i\right)  -\mathcal{O}^{\prime}\!\left(
i\right)  \mathcal{O}\left(  j\right)  \right]  R\left(  j|i\right)  P^{\ast
}\left(  i\right)
\]
which is, on the one hand,%
\begin{equation}
\mathcal{A}\left(  \varepsilon\right)  =\sum_{i,j}\mathcal{O}^{\prime}\!\left(
j\right)  \mathcal{O}\left(  i\right)  \left[  R\left(  j|i\right)  P^{\ast
}\left(  i\right)  -R\left(  i|j\right)  P^{\ast}\left(  j\right)  \right]
=\sum_{i,j}\mathcal{O}^{\prime}\!\left(  j\right)  \mathcal{O}\left(  i\right)
K^{\ast}\left(  j|i\right)  \label{A-K}%
\end{equation}
In this form, it is recognizable as a second moment of 
the distribution\footnote{Since $K^{\ast}$ is defined on a directed link 
(from $i$ to $j$), it can be regarded as a distribution (of probability 
currents) on a directed network on configuration space: 
$ \left\{ \mathcal{C}_{i} \right\} $.  }  $K^{\ast}$.

On the other hand, we may define%
\begin{equation}
\Delta\mathcal{O}^{\prime}\!\left(  i\right)  \equiv\left[  \sum_{j}%
\mathcal{O}^{\prime}\!\left(  j\right)  R\left(  j|i\right)  \right]
-\mathcal{O}^{\prime}\!\left(  i\right)  \label{DeltaO}%
\end{equation}
(and similarly, $\Delta\mathcal{O}\left(  i\right)  $) and see that
$\mathcal{A}$ is also%
\[
\mathcal{A}\left(  \varepsilon\right)  =\sum_{i}\left[  \Delta\mathcal{O}%
^{\prime}\!\left(  i\right)  \mathcal{O}\left(  i\right)  -\mathcal{O}^{\prime
}\left(  i\right)  \Delta\mathcal{O}\left(  i\right)  \right]  P^{\ast}\left(
i\right)
\]
since the extra term in the $\Delta\mathcal{O}$'s cancel. Note that
$\Delta\mathcal{O}$ is recognizable as the average \textit{change} in the
observable $\mathcal{O}$ over one step (associated with each configuration
$\mathcal{C}_{i}$) as a result of the dynamics specified by $R$'s. With the
definition (\ref{DeltaO}), we have incorporated time differences (in a single
step) into what \textit{appears} to be an \textit{equal time} correlation:%
\begin{equation}
\mathcal{A}\left(  \varepsilon\right)  =\left\langle \mathcal{O}%
\Delta\mathcal{O}^{\prime}-\mathcal{O}^{\prime}\Delta\mathcal{O}\right\rangle
^{\ast}\label{A-L}%
\end{equation}
which, in the steady state, does not depend explicitly on time $\tau$ or the
difference $\upsilon$.

To summarize, for any observable $\mathcal{O}\left(  i\right)  $ which is a
functional of the system's configuration ($\mathcal{C}_{i}$ or $i$), we can
associate another: $\Delta\mathcal{O}$. Given by Eqn.(\ref{DeltaO}),
$\Delta\mathcal{O}$ encodes the \textit{change} in $\mathcal{O}$ over a single
time step, averaged over the conditional probabilities specified by the
dynamics ($P\left(  j,\varepsilon|i,0\right)  =R\left(  j|i\right)  $). With
this definition, there are two equivalent ways to represent $\mathcal{A}%
\left(  \varepsilon\right)  $, a simple measure of TRA associated with two
distinct observables $\mathcal{O}$ and $\mathcal{O}^{\prime}$. 
Eqn.(\ref{A-K}) shows its connection to the persistent probability current
$K^{\ast}$, while Eqn.(\ref{A-L}) displays its relationship to an
anti-symmetric combination of the observables and their changes in a single
step $\varepsilon$. The latter will provide a clear motivation for studying
the \textquotedblleft probability angular momentum\textquotedblright\ 
\footnote{This concept was introduced in Ref.\cite{WFMNZ21}.} in the
next subsection.

\subsection{Probability angular momentum $\mathcal{L}_{\alpha\beta}$}

\label{sec:PAM}For most physical systems, the associated configuration space
is sufficiently rich that it is described by many variables: $\left\{
\xi_{\alpha}\right\}  ,\alpha=1,2,...$ In this setting, let us denote
$\xi_{\alpha}$ by $\vec{\xi}$ and write $P\left(  \vec{\xi},\tau\right)  $ in
lieu of the abstract notation $P\left(  j,\tau\right)  $. For convenience, we
will assume they are continuous, though our discussion can be readily
generalized to discrete ones. Also, \textquotedblleft
observables\textquotedblright\ of configurations $\mathcal{C}_{i}$ would be
just functions of $\vec{\xi}$. Thus, the simplest \textquotedblleft
observable\textquotedblright\ $\mathcal{O}$ is just $\xi_{\alpha}$, with its
expectation known as the mean $\left\langle \xi_{\alpha}\right\rangle _{\tau
}\equiv\int\xi_{\alpha}P\left(  \vec{\xi},\tau\right)  d\vec{\xi}$. More
generally, we have higher correlations of \textit{different }$\mathcal{O}$'s
at \textit{unequal} times, e.g., $\left\langle \xi_{\alpha}\xi_{\beta}%
^{\prime}\right\rangle _{\tau,\tau^{\prime}}\equiv\int\xi_{\alpha}\xi_{\beta
}^{\prime}\mathcal{P}\left(  \vec{\xi},\tau\cap\vec{\xi}^{\prime},\tau
^{\prime}\right)  d\vec{\xi}d\vec{\xi}^{\prime}$, where $\mathcal{P}$ is the
\textit{joint probability }to find the system at $\left(  \vec{\xi}%
,\tau\right)  $ \textit{and} $\left(  \vec{\xi}^{\prime},\tau^{\prime}\right)
$. In this formulation, the simplest example for $\mathcal{A}\left(
\varepsilon\right)  $ -- the TRA for\ the steady state in a single step
($\tau-\tau^{\prime}=\varepsilon$) -- is%
\begin{equation}
\left\langle \mathcal{M}_{\alpha\beta}\right\rangle \equiv\int\xi_{\alpha}%
\xi_{\beta}^{\prime}K^{\ast}\left(  \vec{\xi}|\vec{\xi}^{\prime}\right)
d\vec{\xi}d\vec{\xi}^{\prime} \label{M of K}%
\end{equation}
Expressed in the form of Eqn.(\ref{A-K}), we emphasize that $\left\langle
\mathcal{M}\right\rangle $ is the second moment of the distribution 
$K^{\ast}$. Note that this is the lowest moment of $K$ we can probe, 
as $K^{\ast}$ is anti-symmetric in its arguments, by definition.

\subsubsection{Formal definition of $\mathcal{L}$ and contrasts with ordinary
angular momenta}
\label{formal}

To express $\mathcal{M}_{\alpha\beta}$ as an expectation of a single
variable/time, we can exploit Eqn.(\ref{DeltaO}) in the form of 
Eqn.(\ref{A-L}):%
\begin{equation}
\left\langle \mathcal{M}_{\alpha\beta}\right\rangle \equiv\int\left[
\xi_{\alpha}\Delta\xi_{\beta}-\xi_{\beta}\Delta\xi_{\alpha}\right]  P^{\ast
}\left(  \vec{\xi}\right)  d\vec{\xi} \label{M&L}%
\end{equation}
Since it is natural to use \textquotedblleft velocity\textquotedblright\ as a
label for%
\begin{equation}
v_{\alpha}\equiv\frac{\Delta\xi_{\alpha}}{\varepsilon} \label{v}%
\end{equation}
let us coin the term average \textquotedblleft probability angular
momentum\textquotedblright\ for%
\begin{align}
\left\langle \mathcal{L}_{\alpha\beta}\right\rangle  &  \equiv\left\langle
\mathcal{M}_{\alpha\beta}\right\rangle /\varepsilon\\
&  =\left\langle \xi_{\alpha}v_{\beta}-\xi_{\beta}v_{\alpha}\right\rangle
^{\ast} \label{PAM}%
\end{align}
The rationale for this term is partly the analogy with the total angular
momentum of a fluid. Associated with a fluid of mass density $\rho\left(
\vec{x},t\right)  $, the current density would be $\vec{J}=\rho\vec{v}$. If
current loops and vortices are present, we may consider the total angular
momentum $\vec{L}=\int\vec{x}\times\vec{v}\rho~d\vec{x}$. The analog for NESS
would be a system with a time independent $\rho$ as well as a non-vanishing
$\vec{J}$ (e.g., a bucket of water under gravity and rotating about its axis
with constant angular velocity). In this case, both $\vec{\nabla}\cdot\vec{J}$
and $\int\vec{J}$ vanish, so that the value of $\vec{L}$ does not depend on
the choice of the origin (of coordinates, $\vec{x}$). In our case, 
Eqn.(\ref{PAM}) is clearly the generalization of $\vec{L}$ in the setting of a
multi-dimensional $\left\{  \xi_{\alpha}\right\}  $.

Since our loops are associated with the flow of probability density, we
believe it is natural to refer to $\mathcal{L}$ as the \textit{probability
angular momenta}. Apart from having more than 3 components in general,
$\mathcal{L}$ turns out to be much more general than $\vec{L}$ in elementary
physics. In particular, the units of $\vec{L}$ are always\footnote{Here,
$\left[  X\right]  $ denotes the units of the variable $X$. $\mathcal{L}$ has
the units of $\vec{L}/m$, also known as the \textit{specific relative angular
momentum}\ in celestial mechanics.} $\left[  mass\right]  \left[  x\right]
^{2}\left[  t\right]  ^{-1}$. Here, since $\int P=1$ is unitless, the first
factor is unnecessary. However, as Eqn.(\ref{PAM}) involves $\xi_{\alpha}$
and $\xi_{\beta}$, we see that%
\begin{equation}
\left[  \mathcal{L}_{\alpha\beta}\right]  =\left[  \xi_{\alpha}\right]
\left[  \xi_{\beta}\right]  \left[  t\right]  ^{-1} \label{[L]}%
\end{equation}
But, in general, the $\xi$'s may have different units for a complex system --
e.g., temperature, pressure, humidity, etc. for studying the global climate.
Thus, we arrive at the units in (\ref{[L]}). (Note that $\left[  P\left(
\xi_{\alpha}\right)  \right]  =\Pi_{\alpha}\left[  \xi_{\alpha}\right]  ^{-1}%
$, as $\int P\Pi_{\alpha}d\xi_{\alpha}=1$; unlike $\rho\left(  \vec{x}\right)
$ of an ordinary fluid with $\left[  \rho\right]  =\left[  mass\right]
\left[  x\right]  ^{-3}$).

Remarkably, another central quantity in stochastic processes -- the diffusion
coefficients $D_{\alpha\beta}$, i.e., the covariance matrix of the noise -- 
also have the same units as $\mathcal{L}_{\alpha\beta}$! 
Indeed, from the way $D$ appears in say, the Fokker-Planck
equation $\partial_{t}P=\Sigma_{\alpha\beta}\partial^{\alpha}\partial^{\beta
}D_{\alpha\beta}P+...$, we see that $\left[  D_{\alpha\beta}\right]  =\left[
\xi_{\alpha}\right]  \left[  \xi_{\beta}\right]  \left[  t\right]  ^{-1}$. As
will be shown (in Appendix A), 
$\left[  D_{\alpha\beta}\right]  =\left[ \mathcal{L}_{\alpha\beta}\right]  $ 
is \textit{not} a coincidence, as they are
the symmetric and anti-symmetric parts of $\left\langle \xi_{\alpha}%
\partial_{t}\xi_{\beta}\right\rangle ^{\ast}$ . Interestingly, it is possible
to associate, intuitively, the notion of an \textquotedblleft
area\textquotedblright\ in the $\xi_{\alpha}$-$\xi_{\beta}$ plane with both
$D$ and $\mathcal{L}$. For convenience, let us choose the origin to be
$\left(  0,0\right)  $ by letting $\vec{\xi}\rightarrow\vec{\xi}-\left\langle
\vec{\xi}\right\rangle ^{\ast}$. Then in the NESS, our system can be thought
of as a point wandering around $\left(  0,0\right)  $ in this plane (with the
other variables projected out). In analog with Kepler's laws, $\mathcal{L}$
controls the \textquotedblleft area\textquotedblright\ swept out per unit time
by this point, on the average. Note that there is a sign associated with the
sweeping movements, \textquotedblleft clockwise\textquotedblright\ being
different from \textquotedblleft anti-clockwise\textquotedblright, so that
cancellation may lead to $\left\langle \mathcal{L}\right\rangle =0$. If that
occurs, we would label our system to be in thermal equilibrium rather than a
NESS. On the other hand, if there are no deterministic forces and the system
is driven by noise alone, its $\left(  \xi_{\alpha},\xi_{\beta}\right)  $ will
perform a random walk in the plane. Starting from $\left(  0,0\right)  $, say,
the ensemble of wanderings (the \textquotedblleft spaghetti\textquotedblright ) 
will appear to cover an ellipse, the area of\ which increases
$\propto$ $t$, with a rate encoded in $D$. In this sense, both $D$ and
$\mathcal{L}$ can be associated with \textquotedblleft an area per unit
time\textquotedblright\ and carry units of $\left[  \xi_{\alpha}\right]
\left[  \xi_{\beta}\right]  \left[  t\right]  ^{-1}$.

\subsubsection{Distribution of $\mathcal{L}$}

While in theory a non-zero value of $\left\langle \mathcal{L}\right\rangle $
signals the presence of TRA, no practical measurements of $\left\langle
\mathcal{L}\right\rangle $ is likely to be exactly zero, even for systems in
equilibrium. To decide if such an average is statistically consistent with
zero or not, a typical means is to compare it to the standard deviation,
$\sigma_{\mathcal{L}}\equiv\left[  \left\langle \mathcal{L}^{2}\right\rangle
-\left\langle \mathcal{L}\right\rangle ^{2}\right]  ^{1/2}$. Unfortunately,
since we are considering general stochastic processes, the noise typically
contribute significantly to $\sigma_{\mathcal{L}}$. Indeed, as will be shown
below, there are many explicit, solvable examples of NESS with $\left\langle
\mathcal{L}\right\rangle \ll\sigma_{\mathcal{L}}$. Thus, demanding
$\left\langle \mathcal{L}\right\rangle \gtrsim\sigma_{\mathcal{L}}$ is too
severe a criterion to distinguish a NESS from equilibrium stationary states.
Instead, we discover that a better guide is to study\footnote{In this 
article, we will restrict our attention for $Q$ only to that in a steady state. 
To be consistent with notation, we should use $Q^{\ast}$, as $Q$, like $P$,
should denote a time dependent quantity. However, for simplicity, we will 
drop the $^{\ast}$ and all references to $Q$ or its Fourier transform 
$\tilde{Q}$ will mean distributions in stationary states.}
$Q\left(  \mathcal{L}\right)  $, the full distribution of $\mathcal{L}$.

To provide another motivation for studying $Q$, note that, if a single
stochastic system is observed in a stationary state (to estimate $\left\langle
\mathcal{L}\right\rangle $ by computing the time average $\mathcal{\bar{L}}$),
$\mathcal{L}$ would typically assume both positive and negative values. To
determine if our system is evolving under detailed balance or not, we would
need to check if a positive value of $\mathcal{L}$\ appears as often as a
negative one. In other words, we can build a histogram from the observed
values of $\mathcal{L}$ and check if it is symmetric or not. Specifically, we
can see if the frequency for $\mathcal{L}$ to be seen in any interval (e.g.,
a bin) $\left[  -\ell,-\ell^{\prime}\right]  $ is statistically the same as in
$\left[  \ell^{\prime},\ell\right]  $. In the next subsection, we will propose
a specific criterion along these lines.

To study $Q$ theoretically, we define (for a specific pair $\alpha,\beta$)%
\begin{equation}
Q\left(  \mathcal{L}\right)  \equiv\left\langle \delta\left[  \mathcal{L}%
-\left(  \xi_{\alpha}v_{\beta}-\xi_{\beta}v_{\alpha}\right)  \right]
\right\rangle ^{\ast} \label{Q}%
\end{equation}
so that $\left\langle \mathcal{L}_{\alpha\beta}\right\rangle =\int%
\mathcal{L}Q\left(  \mathcal{L}\right)  d\mathcal{L}$. Naturally, we can
expect computing $Q\left(  \mathcal{L}\right)  $ to be challenging in general.
However, its Fourier transform
\begin{equation}
\tilde{Q}\left(  z\right)  \equiv\int e^{iz\mathcal{L}}Q\left(  \mathcal{L}%
\right)  d\mathcal{L=}\left\langle \exp iz\left(  \xi_{\alpha}v_{\beta}%
-\xi_{\beta}v_{\alpha}\right)  \right\rangle ^{\ast} \label{Q-ft}%
\end{equation}
may be more accessible. In particular, a closed form for $\tilde{Q}$ can be
obtained for \textquotedblleft linear Gaussian models,\textquotedblright\ as
will be shown below. Analyzing its singularities (in complex $z$) will provide
some insight into the asymmetry: $Q\left(  \mathcal{L}\right)  $ \textit{vs.}
$Q\left(  -\mathcal{L}\right)  $, especially for large $\mathcal{L}$. In all
cases, since $Q$ can also be represented by $\int\delta\left[  \mathcal{L}%
-\left(  \xi_{\alpha}\xi_{\beta}^{\prime}-\xi_{\beta}\xi_{\alpha}^{\prime
}\right)  \right]  R\left(  \vec{\xi}^{\prime}|\vec{\xi}\right)  P^{\ast
}\left(  \vec{\xi}\right)  $, we readily verify that $Q\left(  \mathcal{L}%
\right)  =Q\left(  -\mathcal{L}\right)  $ when DB is satisfied, i.e., when
$R\left(  \vec{\xi}^{\prime}|\vec{\xi}\right)  P^{\ast}\left(  \vec{\xi
}\right)  =R\left(  \vec{\xi}|\vec{\xi}^{\prime}\right)  P^{\ast}\left(
\vec{\xi}^{\prime}\right)  $.

\subsubsection{Measuring $\mathcal{L}$ from a time series}
\label{timeSeries}
The results presented above may appear theoretical and formal. Let us turn to
ways they can be implemented in practice. First, predictions for $\left\langle
\mathcal{L}\right\rangle $ are based on averages over statistical ensembles
such as $P^{\ast}\left(  \xi\right)  $, which are typically, not readily
available. Instead, real data are collected as one or more time series, on
which statistical analyses are performed. To distinguish the theoretical
$\mathcal{L}$ from a similar quantity in practice, we denote the latter by
$\ell$. In the end, we will be comparing, e.g., $\left\langle \mathcal{L}%
\right\rangle $ with averages of $\ell$.

Generally, observations of a system (physical or in computer simulations)
consist of the time series of a number of variables. In principle, to find
averages using $P\left(  \vec{\xi},\tau\right)  $, we need time series from an
ensemble of \textquotedblleft identical\textquotedblright\ systems (ideally,
an infinite number of them). In practice, a rough estimate of $P$ are often
made from the history of a few systems (spaghetti plots). However, if we focus
our interest on \textit{steady states}, we can take a single, very long series
and (assuming the system is sufficiently ergodic over the span of the
observations) replace ensemble averages $\left\langle \cdot\right\rangle
^{\ast}$ by time averages, $\overline{{\Large \cdot}}$ . See (\ref{ell2L})
below as an example.

Proceeding along these lines, let us denote a long time series of the
variables of interest by $\xi_{\gamma}\left(  \tau\right)  ;$ $\tau
=0,\varepsilon,2\varepsilon,...,M\varepsilon$. To detect TRA in the form of
probability angular momenta, we can take any pair of variables (or two
different linear combinations of $\xi$'s) and form a third time 
series\footnote{Note that this $\ell$ is identical to, but slightly simpler
than, the version with $\xi\left(  \tau\right)  $ and $v\left(  \tau\right)
=\Delta\xi\left(  \tau+\varepsilon\right)  /\varepsilon=\left[  \xi\left(
\tau+\varepsilon\right)  -\xi\left(  \tau\right)  \right]  /\varepsilon$.}:
\[
\ell_{\alpha\beta}\left(  \tau\right)  =\frac{1}{\varepsilon}\left[
\xi_{\alpha}\left(  \tau\right)  \xi_{\beta}\left(  \tau+\varepsilon\right)
-\xi_{\beta}\left(  \tau\right)  \xi_{\alpha}\left(  \tau+\varepsilon\right)
\right]
\]
Clearly, it is not significant that this series has one less element than in
the original set. Averaging $\ell$ over the run provides us with a quantity%
\[
\overline{\ell_{\alpha\beta}}\equiv\frac{1}{M}\sum_{\tau=0}^{M-1}\ell
_{\alpha\beta}\left(  \tau\right)
\]
which we can identify with the average probability angular momentum
(\ref{M&L}-\ref{PAM}):%
\begin{equation}
\overline{\ell_{\alpha\beta}}\rightarrow\left\langle \mathcal{L}_{\alpha\beta
}\right\rangle \label{ell2L}%
\end{equation}

Given $N$ variables, there are $N\left(  N-1\right)  /2$ such $\bar{\ell}$'s.
If \textit{any }of them is non-zero, we would conclude that the system is in a
NESS rather than in thermal equilibrium. However, to apply such a statement to
any given situation is not straightforward, since there are always statistical
uncertainties. Even when we are working with a system that satisfies DB (and
so, theoretically TR symmetric), any particular computational result for
$\bar{\ell}$ is unlikely to be identically zero. One gauge for whether the
average of some quantity can be considered \textquotedblleft statistically
zero\textquotedblright\ is to compare it to the standard deviation
$\sigma_{\ell}\equiv [ \overline{\ell^{2}}-\bar{\ell}^{2} ]  ^{1/2}%
$. However, as noted earlier and illustrated below, there are many systems
which displays $\bar{\ell}\ll\sigma_{\ell}$ despite being driven by a known,
DB violating, dynamics. The issue here is that although DB is either satisfied
or not, the signals of DB violation can be \textit{arbitrarily small}. In
other words, a system can be \textquotedblleft infinitesimally
near\textquotedblright\ thermal equilibrium (or\ \textquotedblleft very far
from equilibrium\textquotedblright). For this reason, we use the label
\textquotedblleft subtle\textquotedblright\ to describe a NESS for which the
signal of TRA are so small that $\bar{\ell}\ll\sigma_{\ell}$.\ 

Detecting such weak signals would require a more sensitive criterion 
than $\bar{\ell}\neq0$. As
proposed for $Q$ above, we turn to the asymmetry of the full distribution of
$\ell$. Specifically, let us use bins of width $w$ centered symmetrically 
around $\ell=0$ : at $\ell_{\pm b}=\pm\left(  b-1/2\right) w, ~b=1,2,...$. 
Denoting the frequency of occurrence in each bin as $H_{\pm b}$, we can 
check for systematic difference between $H_{b}$ and $H_{-b}$. Then, 
estimating statistical fluctuations by the square roots of the absolute
numbers within the bins (e.g., $\left[  H_{b}+H_{-b}\right]  ^{1/2}$), we can
get a better sense of whether TRA is present or not. Thus, we propose to study
\begin{equation}
\Upsilon_{b}\equiv\frac{H_{b}-H_{-b}}{\left[  H_{b}+H_{-b}\right]  ^{1/2}}
\label{Omega_b}%
\end{equation}
and check if it lies \textit{systematically} higher than $+1$, or lower than
$-1$. Below, we will illustrate this process with times series from simulation
results based on the linear Gaussian model (LGM, used extensively in climate
modeling \cite{Penland93,WFMNZ21}, and suitable for coupled simple harmonic
oscillators in contact with two thermal baths). Since LGMs are solvable
analytically, those results can be used as a guide for how well these methods work.

\subsection{Asymmetries for the time series of a single variable}

\label{sec:1D}The previous subsection may give the (wrong) impression that
persistent current loops are absent from systems in \textquotedblleft one
dimension.\textquotedblright\ After all, in elementary physics, there can be
no angular momentum for motion in 1D! The correct statement is the following:
If the rates $R\left(  i|j\right)  $ are non-trivial only for a \textit{open
chain} of configurations, $\mathcal{C}_{1},...,\mathcal{C}_{N}$ (i.e.,
$R\left(  i|j\right)  \equiv0$ for all $i>j\pm1$), then $K^{\ast}$ must be
identically zero. Note that, in the limit of continuous $\mathcal{C}$ space
(say, $x$), the Master equation for such systems becomes a Fokker-Planck
equation in 1D: $\partial_{t}P\left(  x,t\right)  =D\partial_{x}^{2}%
P-\partial_{x}\left(  \mu P\right)  $ on an \textit{open} interval. Thus,
$\partial_{t}P^{\ast}=0$ implies that the steady state probability current,
$\mu P^{\ast}-D\partial_{x}P^{\ast}$, must vanish\footnote{Indeed, for any
$\mu\left(  x\right)  $ on an open interval, we can integrate it to $V\left(
x\right)  $, specifically $\mu=-\partial_{x}V$. Then we easily obtain
$P^{\ast}\left(  x\right)  \propto-V\left(  x\right)  /D$ and $J_{FP}^{\ast
}=0$.}. As soon as we allow jumps further than \textquotedblleft nearest
neighboring $\mathcal{C}$'s,\textquotedblright\ it is possible to have DB
violation and steady current loops, even in 1D. Then, TRA can be detected.

To phrase these considerations another way, if we are provided with the time
series of a \textit{single} stochastic variable -- denoted by $\zeta\left(
\tau\right)  $ (may be continuous or discrete) -- in a stationary state, it is
still possible to determine if there is any statistical asymmetry between a
forwards version of that \textquotedblleft movie\textquotedblright\ (time
series) and a backwards one. This line of inquiry was motivated by a recent
work of Mori, Majumdar and Schehr\cite{MMS}. To detect NESS, these authors
exploited the location of an extremal value of a single variable -- within a
finite interval in the time series. Our interpretation of this approach is the
following. What enters effectively is a \textit{three-point} correlation
function (3pf), with two of the times being the ends of the interval and the
third time being at the extremal event. In this context, a simpler observable
can be used, e.g., $\Theta$ below\footnote{Like $Q$ above, $\Theta$ 
should also have a superscript *, but we will drop that here. Similarly, we 
will drop * when we write $\Gamma$ and $\tilde{\Gamma}$ below.}. 
Further, in a certain limit, such a 3pf
becomes the correlation of two \textit{different} observables (functions of
$\zeta$) at just \textit{two} times, a principle which underlies Eqns.(\ref{2-2},
\ref{22TRA}).

We begin with the conceptually simpler, generic 3pf in a stationary state:%
\begin{equation}
\Theta\left(  \tau,\upsilon\right)  \equiv\left\langle \zeta\left(  0\right)
\zeta\left(  \tau\right)  \zeta\left(  \tau+\upsilon\right)  \right\rangle
^{\ast} \label{3pf}%
\end{equation}
Here, we have exploited time translation invariance in the NESS to set the
argument of the first time to be $0$. To be precise, this 3pf is, using the
notation of joint probabilities at \textit{three} times:
\[
\int\xi\eta\zeta\mathcal{P}^{\ast}\left(  \xi,\tau+\upsilon\cap\eta,\tau\cap
\zeta,0\right)  d\xi d\eta d\zeta
\]
More explicitly, it is given by%
\[
\int\xi\eta\zeta P\left(  \xi,\upsilon|\eta,0\right)  P\left(  \eta,\tau
|\zeta,0\right)  P^{\ast}\left(  \zeta\right)  d\xi d\eta d\zeta
\]
where $\tau,\upsilon>0$ and $P\left(  \zeta,\tau|\eta,0\right)  $ is the
conditional probability discussed above. Since $\zeta$ is a single (commuting)
variable, there is no loss of generality to focus on just the first quadrant:
$\tau,\upsilon>0$. For example, $\Theta\left(  -1,-2\right)  =\Theta\left(
2,1\right)  $, because time translation means $\left\langle \zeta\left(
0\right)  \zeta\left(  -1\right)  \zeta\left(  -3\right)  \right\rangle
^{\ast}=\left\langle \zeta\left(  3\right)  \zeta\left(  2\right)
\zeta\left(  0\right)  \right\rangle ^{\ast}$, which is the same as
$\left\langle \zeta\left(  0\right)  \zeta\left(  2\right)  \zeta\left(
3\right)  \right\rangle ^{\ast}$.

In this setting, TRA would manifest as%
\[
\Theta\left(  \tau,\upsilon\right)  \neq\Theta\left(  \upsilon,\tau\right)
\]
Note that this inequality cannot occur if $\upsilon=\tau$, a case
corresponding to the correlation of three points symmetrically distributed in
time (e.g., $\left\langle \zeta\left(  0\right)  \zeta\left(  1\right)
\zeta\left(  2\right)  \right\rangle ^{\ast}=\left\langle \zeta\left(
-1\right)  \zeta\left(  0\right)  \zeta\left(  1\right)  \right\rangle ^{\ast
}$). Conversely, if the three $\zeta$'s are asymmetrically located
($\upsilon\neq\tau$), then $\Theta\left(  \tau,\upsilon\right)  -\Theta\left(
\upsilon,\tau\right)  $ does not necessarily vanish.

From the above discussion, there is no other restriction on $\upsilon,\tau$.
In particular, it is possible to consider the limit of one or the other being
zero. Then, we are faced with, say,
\begin{align*}
\Gamma\left(  \upsilon\right)   &  \equiv\left\langle \zeta^{2}\left(
0\right)  \zeta\left(  \upsilon\right)  \right\rangle ^{\ast}\\
&  =\int\eta\zeta^{2}P\left(  \eta,\upsilon|\zeta,0\right)  P^{\ast}\left(
\zeta\right)
\end{align*}
and a measure for TRA in the system is $\Gamma\left(  \upsilon\right)
-\Gamma\left(  -\upsilon\right)  $, i.e.,%
\begin{align}
\tilde{\Gamma}\left(  \upsilon\right)   &  \equiv\left\langle \zeta^{2}\left(
0\right)  \zeta\left(  \upsilon\right)  \right\rangle ^{\ast}-\left\langle
\zeta\left(  0\right)  \zeta^{2}\left(  \upsilon\right)  \right\rangle ^{\ast
}\label{Gamma-tilde}\\
&  =\int\eta\zeta\left(  \zeta-\eta\right)  P\left(  \eta,\upsilon
|\zeta,0\right)  P^{\ast}\left(  \zeta\right) \nonumber
\end{align}
Since we fully expect decorrelation at large $\upsilon$ (most likely
exponentially decaying, so that $\Gamma\left(  \upsilon\right)  \rightarrow
\left\langle \zeta\right\rangle ^{\ast} 
\left\langle \zeta^{2}\right\rangle ^{\ast}$),
we may consider summing over all $\upsilon$ to define the total asymmetry
$\tilde{\Gamma}_{tot}\equiv\sum_{\upsilon>0}\tilde{\Gamma}\left(
\upsilon\right)$
If the asymmetry does not oscillate with $\tau$, this quantity should provide
a sizable signal for TRA. To be brief here, we will postpone this line of
inquiry to a future study. 

As in the previous subsection, we will next assume that ensemble averages can
be replaced by time averages. Consequently, given the time series of a single
variable in a \textit{steady state}, $\zeta\left(  \tau\right)  $, we can
construct the following (average over $\tau$) for measuring $\tilde{\Gamma
}\left(  \upsilon\right)  $ when a time series is given:%
\[
\overline{\zeta\left(  \tau\right)  \zeta\left(  \tau+\upsilon\right)  \left[
\zeta\left(  \tau+\upsilon\right)  -\zeta\left(  \tau\right)  \right]  }%
\]
An example of $\tilde{\Gamma}\left(  \upsilon\right)  $ in a simple system 
will be provided below.  Also analogous to extending our study of 
$\left\langle \mathcal{L} \right\rangle $ to $Q\left(  \mathcal{L}\right)  $, 
we can consider
distributions of $\tilde{\Gamma}$, a line of research which may be pursued in
the future. Here, let us end with a simpler means to gauge if a given value of
$\tilde{\Gamma}$ can be regarded as significant or not. To set a scale for
this, we can introduce a dimensionless ratio:%
\[
\tilde{\Gamma}/\sigma_{\zeta}^{3}%
\]
where $\sigma_{\zeta}\equiv\left[  \left\langle \zeta^{2}\right\rangle
-\left\langle \zeta\right\rangle ^{2}\right]  ^{1/2}$ is the standard
deviation in $\zeta$. Such a measure is consistent with the standard used for
skewness, (also a three point correlation).

To recapitulate, we have provided in this section a general framework for
detecting detailed balance violation in non-equilibrium steady states. The
unifying, underlying theme is persistent probability currents ($K^{\ast}$ or
$\vec{J}_{FP}^{\ast}$) and loops. Though this may seem to be an abstract
concept, the currents do manifest themselves in concrete, \textquotedblleft
observable\textquotedblright\ quantities.\ In particular, the main consequence
of DB violation and non-trivial $K^{\ast}$ is time reversal asymmetry. To
highlight these, we proposed several simple ways. For systems with many
variables, probability angular momenta (and their distributions) represent the
simplest possibilities, as they involve just two-point correlations at unequal
times (i.e., second moments of $K^{\ast}$ or $\mathcal{P}^{\ast}$). For systems 
with a single variable, however, the simplest possibility involves three-point
functions or correlations of higher powers at two times (i.e., third moments).
In the rest of this article, many examples of these asymmetries will be presented, 
from exactly solvable \textquotedblleft toy\textquotedblright\ models to 
complex physical phenomena.

\section{Examples of time reversal asymmetry: Subtle, manifest, and
extraordinary}
\label{Examples}

In the previous section, generalities were provided. The following
subsections will be devoted to specific examples, illustrating how we can
detect various signals of DB violation and TRA in NESS, from the well hidden
and the manifestly obvious to the completely unexpected. Specifically, much of
our studies will involve probability angular momenta, as they are the most
direct consequences of persistent probability current loops. In general, given
a set of DB violating $R$'s, the stationary distribution $P^{\ast}$ is not
explicitly known (unlike the well known Boltzmann factor for systems in
thermal equilibrium). Therefore, we believe it is helpful to begin with some
simple and exactly solvable models (i.e., those with exactly known
$P^{\ast}$), so that the concepts introduced above can be illustrated in some
detail. In a final subsection, we turn to some recent discoveries in a simple
driven diffusive lattice gas, which displays various surprising behaviors --
inconceivable for systems in thermal equilibrium with maximal entropy.

\subsection{Simple, exactly solvable systems}

In this subsection, we present four very simple systems, all being exactly
solvable. These should provide readers unfamiliar with these ideas an easy
introduction into the workings of DB violation, TRA, current loops and
probability angular momenta in NESS. The first toy model, with just three
configurations, cannot be simpler. The second system is motivated more by
physics: coupled simple harmonic oscillators in contact with two thermal
baths. The third is a generalization to arbitrary numbers of degrees of
freedom, driven linearly and subjected to additive white noise. Finally, we
consider a biased random walker on a half-line, with reset to the origin, so
that its configuration space is one-dimensional.

\subsubsection{Simple three-configuration system coupled to two thermal
reservoirs}

\label{3st}If a system has just two configurations (i.e., micro-states), then
there can be no transition loops. The simplest one which can support loops is
a 3-configuration system, in which there is a single loop. Not surprisingly,
such a system is easily solvable and can serve to illustrate the ideas
outlined in the Sections above.

Let us label the three configurations by $i$ ($=1,2,3$), which we may imagine
to be levels with energy $\left(  i-1\right)  E$. The most general set of
transition rates would involve six quantities, but let us focus on a very
simple NESS, by coupling the system to two reservoirs, at temperatures $T$ and
$T^{\prime}$. To further simplify matters, using discrete time ($\tau$), we
couple these reservoirs to the transitions which change energy by $E,2E$ and
exploit Metropolis rates: $\min\left\{  1,\varphi\right\}  $ and $\min\left\{
1,\tilde{\varphi}\right\}  $, where $\varphi\equiv e^{-E/k_{B}T}$ and
$\tilde{\varphi}\equiv e^{-2E/k_{B}T^{\prime}}$. Obviously, DB is retrieved if
$\tilde{\varphi}=\varphi^{2}$ and the system will settle into thermal
equilibrium at temperature $T$.

With this setup, the Master equation for $\left\vert P\right\rangle $ (i.e.,
$P\left(  i,\tau\right)  $, the probability for finding the system in $i$ and
evolving in $\tau$) is $\Delta_{\tau}\left\vert P\right\rangle =\mathfrak{L}%
\left\vert P\right\rangle $. Deferring details (of $\mathfrak{L}$ and its
eigenvalues) to Appendix B, we display only the steady state distribution here:%

\[
\left\vert P^{\ast}\right\rangle =\frac{1}{Z}\left(
\begin{array}
[c]{c}%
2+\varphi\\
2\varphi+\tilde{\varphi}\\
\varphi^{2}+\tilde{\varphi}\varphi+\tilde{\varphi}%
\end{array}
\right)
\]
where $Z=\left(  2+\varphi\right)  \left(  1+\varphi+\tilde{\varphi}\right)
$. Note that, when $T=T^{\prime}$, and DB is satisfied, we have $\tilde
{\varphi}=\varphi^{2}$ and retrieve the usual Boltzmann factors of thermal
equilibrium: $\left(  1,\varphi,\varphi^{2}\right)  /\left(  1+\varphi
+\varphi^{2}\right)  $. In a NESS, the net current from $i=1$ to $2$ is%
\[
K^{\ast}\left(  2|1\right)  =\varphi P^{\ast}(1)-P^{\ast}(2)=\left(
\varphi^{2}-\tilde{\varphi}\right)  /Z
\]
(and vanishes if $T=T^{\prime}$, as expected). With three states, there is
just a single loop and we can verify that the other two net currents also
assume this value. Note that sign of the current loop ($1\rightarrow
2\rightarrow3\rightarrow1$) is indicative of the relative \textquotedblleft
push\textquotedblright\ from the two reservoirs. If $T>T^{\prime}$, then we
expect the former to \textquotedblleft drive\textquotedblright\ the system and
so, the transitions $1\rightarrow2\rightarrow3$ will be more prevalent than
$1\rightarrow3$. The extreme case of $T^{\prime}=0$ is interesting, as such a
reservoir only drains energy (from $3\rightarrow1$) and would ordinarily
prevent level $3$ from being occupied. By coupling the system to two
reservoirs, this level can maintain a non-trivial occupation, provided $T>0$.
As for angular momentum, it is rather pointless (though possible) to assign a
2D space to 3 states and compute a single value, since the single value of
$K^{\ast}$ carries all the information about this NESS.

\subsubsection{Coupled simple harmonic oscillators (SHO) in contact with two
heat baths}

\label{sec:TTHSO}Another exactly solvable system evolving with DB violating
dynamics is the \textquotedblleft two-temperature simple harmonic
oscillators\textquotedblright\ (TTSHO). Here, we consider the simplest case,
with just two SHOs (each in 1D with displacements $x_{1}$ and $x_{2}$, denoted as
$\vec{x}$) coupled to thermal baths at temperatures $T_{1}$ and $T_{2}$.
Physically, this system is the massless limit of $\vec{F}=m\vec{a}$, in which
the force consists of a deterministic part ($-\vec{\nabla}V\left(  x_{1,}%
x_{2}\right)  $) and terms associated with the thermal reservoirs: damping
($-\lambda\partial_{t}\vec{x}$) and noise ($\vec{\eta}$). Specifically, we use
the usual quadratic potential (with spring constants $k$ and $k_{\times}$)%
\[
V\left(  \vec{x}\right)  =\frac{1}{2}\left\{  k\vec{x}^{2}+k_{\times}\left(
x_{1}-x_{2}\right)  ^{2}\right\}
\]
so that the Langevin equation can be written as $\partial_{t}\vec{x}%
=-\vec{\nabla}V+\vec{\eta}\equiv-\mathbb{F}\vec{x}+\vec{\eta}$. (after
absorbing $\lambda$ into $t$). The additive Gaussian noise obeys $\left\langle
\vec{\eta}\right\rangle =0$ and correlation $\left\langle \vec{\eta}\left(
t\right) \vec{\eta} \left(  t^{\prime}\right) ^T \right\rangle
=2\mathbb{D}\delta\left(  t-t^{\prime}\right)  $. Here, $\mathbb{D}$ and
$\mathbb{F}$ are matrices:%
\[
\mathbb{D}=\left(
\begin{array}
[c]{cc}%
T_{1} & 0\\
0 & T_{2}%
\end{array}
\right)  ,~~\mathbb{F}=\left(
\begin{array}
[c]{cc}%
k+k_{\times} & -k_{\times}\\
-k_{\times} & k+k_{\times}%
\end{array}
\right)
\]
The associated Fokker-Planck equation is $\partial_{t}P=\vec{\nabla}%
\cdot\left\{  \mathbb{D}\vec{\nabla}P+\mathbb{F}\vec{x}P\right\}  $. The
physics of $\mathbb{F}$ is elementary: a slow restoring force associated with
the oscillators moving in phase, and a fast mode moving $180^{\circ}$ out of
phase. If the noise on both are the same (i.e., $T_{1}=T_{2}=T$), then the
system will settle into the standard Boltzmann stationary $P^{\ast}$
($\propto\exp\left(  -V/T\right)  $ with $k_{B}\equiv1$). Otherwise, if
$k_{\times}$ is absent, $P^{\ast}$ will be the product of two Boltzmann
factors, one for each $x_{\alpha}$ and $T_{\alpha}$.

However, if neither $k_{\times}$ nor $T_{1}-T_{2}$ vanishes, then DB is
violated, but $P^{\ast}$ will still be a
Gaussian\cite{MLax60,UCT98,JBW03,ZS07}. Indeed, we can easily show that
$P^{\ast}\propto\exp\left(  -\vec{x}\cdot\mathbb{C}^{-1}\vec{x}/2\right)  $
satisfies $\partial_{t}P^{\ast}=0$, with $\mathbb{C}$ being the solution to
the linear equation\cite{JBW03}%
\begin{equation}
\mathbb{FC+CF}^{T}=2\mathbb{D} \label{FC+CF}%
\end{equation}
The proof relies on casting (the Fokker-Planck version of) the probability
current
\[
\vec{J}_{FP}^{\ast}=-\left[  \mathbb{D}\vec{\nabla}+\mathbb{F}\vec{x}\right]
P^{\ast}=\left(  -\mathbb{D}+\mathbb{FC}\right)  \vec{\nabla}P^{\ast}%
\]
as $\vec{J}_{FP}^{\ast}=\mathbb{A}\vec{\nabla}P^{\ast}$ with%
\[
\mathbb{A}=\frac{1}{2}\left[  \mathbb{FC}-\mathbb{CF}^{T}\right]
\]
Since $\mathbb{A}$ is anti-symmetric, we have $-\partial_{t}P^{\ast}%
=\vec{\nabla}\cdot\vec{J}_{FP}^{\ast}=\vec{\nabla}\cdot\mathbb{A}\vec{\nabla
}P^{\ast}=0$.

As a $2\times2$ matrix, the only independent component of $\mathbb{A}$
is\cite{UCT98}

\begin{equation}
A_{12}=\frac{1}{2}\frac{k_{\times}}{k+k_{\times}}\left(  T_{1}-T_{2}\right)
\label{L12}%
\end{equation}
Next, we readily find the probability angular momentum 
(from $\int\vec{x}\times \vec{J}_{FP}^{\ast}$) to be 
\[
\left\langle \mathcal{L}_{12}\right\rangle ^{\ast}
= \int\left[  x_{1}A_{21}\partial_{1}-x_{2}A_{12}%
\partial_{2}\right]  P^{\ast}d\vec{x}=2A_{12}
\]
Note that, as expected, it vanishes if $T_{1}=T_{2}$ or 
$k_{\times}=0$.

\begin{figure}[pth]
\centering
\includegraphics[width=0.85\textwidth]{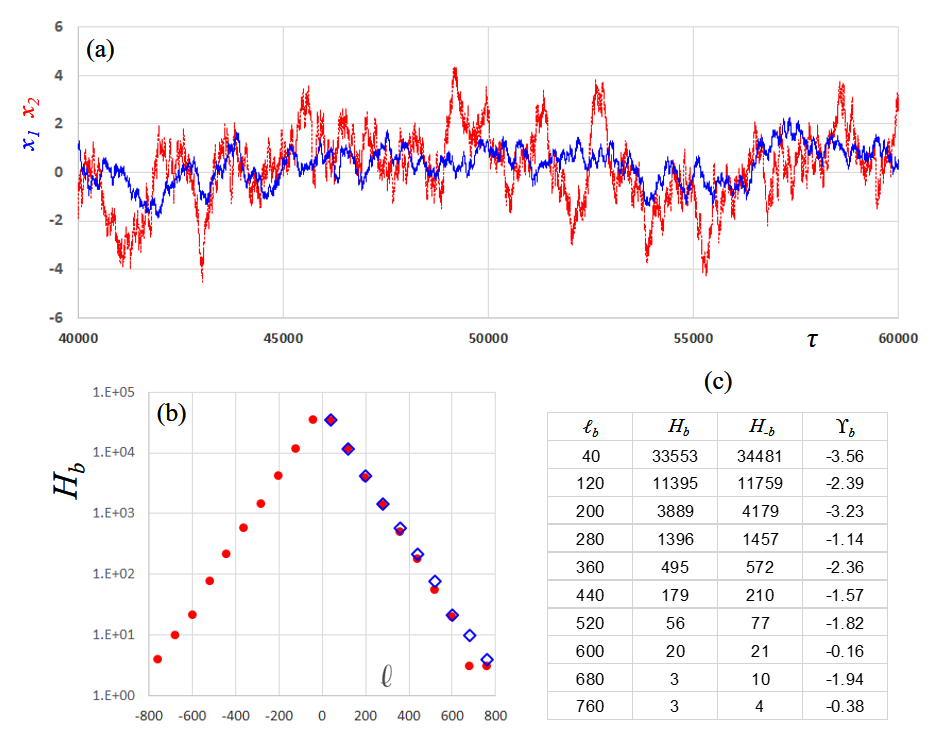}\caption{ Simulation of
coupled SHO's in contact with two thermal baths. (a) A section of the time
series (blue/red in contact with colder/hotter bath). (b) Histogram of $\ell$ 
($H_{b}$). Blue diamonds are frequencies for bins with $\ell<0$, showing
systematically higher values. (c) Table of center of bins (value of $\ell$)
and frequencies $H_{\pm b}$, as well as the ratio $\Upsilon_{b}$ (see text for
details). }%
\label{fig:TTSHO}%
\end{figure}

To illustrate the TTSHO, we simulate the Langevin equation by Euler's method
of evolution, discretizing $t$ by steps of $\varepsilon\ll1$: $\vec{x}\left(
t+\varepsilon\right)  =\vec{x}\left(  t\right)  +\varepsilon\left\{
-\mathbb{F}\vec{x}\left(  t\right)  +\vec{\eta}\left(  t\right)  \right\}  $,
with Gaussian noise of co-variance\footnote{$\delta$ denotes the Kronecker delta here.} 
$2\mathbb{D}\delta\left(  t,t^{\prime}\right) /\varepsilon$. Fig.
\ref{fig:TTSHO} shows a section of the time trace of this system in a
stationary state -- with $k=2=2k_{\times}$, $\left(  T_{1},T_{2}\right)
=\left(  1,7\right)  $, $\varepsilon=0.001$. From this trace, it is far from
clear that TRA is present and the state is a NESS. When plotted in the $x_{1}%
$-$x_{2}$ plane, the trajectory also does not display obvious sense of
rotation, clockwise or anticlockwise. A better test is by measuring the
probability angular momentum (averaged over $100K$ steps of the run) 
$\bar{\ell}$, which is $\simeq-1.923$. Though it is acceptably close to the
theoretical $-2.0$, the issue is: If we had only the time series and not the
theory, would we be able to conclude that this $\bar{\ell}$ is not
\textquotedblleft statistically zero\textquotedblright? If we naively compare
it to the observed standard deviation ($\sigma_{\ell}\simeq107.62$), we find
$\sigma_{\ell}$ to be over 50 times $\bar{\ell}$ ! To be more confident,
we consider $H_{b}$, the histogram associated with $\ell$. The observed $\ell$
values\footnote{Note that typical $x$'s are $O\left(  1\right)  $ here
(comparable to $\mathbb{F}$). But, the typical velocities ($\partial x$'s) are
$O\left(  100\right)  $ (controlled by $\eta\sim\sqrt{\mathbb{D}/\varepsilon}%
$), so that many $\ell$'s are also $O\left(  100\right)  $.} range as far as
$\pm800$. Illustrating with 20 bins ($w=80$, centered at $\ell_{\pm b}=$
$\pm40,\pm120,...$) in Fig.\ref{fig:TTSHO}b, we see that the counts for the
$\ell<0$ bins are systematically larger than those in the $\ell>0$ ones. In
Fig.\ref{fig:TTSHO}c, we show the bin centers $\ell_{\pm b}$ and the
frequencies $H_{\pm b}$. More crucially, we see that every $\Upsilon_{b}%
$\ (from Eqn.\ref{Omega_b}) is fairly negative (apart from two cases with
$\left\vert \Upsilon_{b}\right\vert <1$, both associated with low counts).
Such plots provide us with confidence that, despite $\bar{\ell}$ being
$\sim\sigma_{\ell}/50$, TRA is indeed present and this system is definitely in
a NESS (as expected from the DB violating rules used to simulate the
stationary state).

\begin{figure}[pth]
\centering
\includegraphics[width=0.45\textwidth]{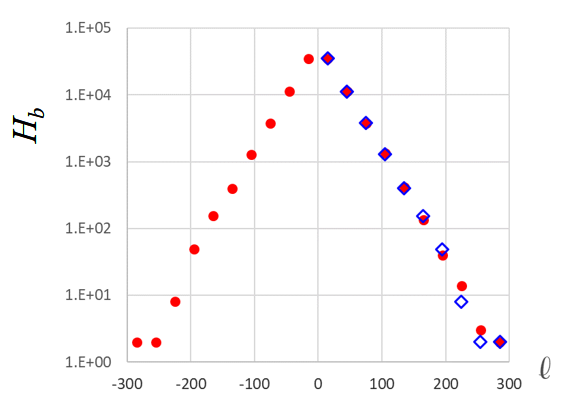}\caption{ Histogram of $\ell$
($H_{b}$) for a simulation of coupled SHO's in contact with a \textit{single}
thermal bath. Blue diamonds are frequencies for bins with $\ell<0$, showing no
systematic asymmetry. }%
\label{fig:SHO}%
\end{figure}

As a contrast,\ we performed simulations for the same system coupled to a
single thermal bath at $T=1$. Here, $\bar{\ell}\simeq-0.0155$ is considerably
smaller, in comparison with $\sigma_{\ell}\simeq38.4$. Meanwhile, 
Fig\ref{fig:SHO} clearly shows there is no \textit{systematic} asymmetry in the
histogram. In addition to the signs of $\Upsilon$ being evenly distributed
around zero, 70\% of the $\left\vert \Upsilon\right\vert $'s are less than
unity. We believe these two cases illustrate the import role played by the
distribution of probability angular momentum for distinguishing whether a
system is in a NESS or in thermal equilibrium (or so close that TRA is 
imperceptible).

\subsubsection{Linear Gaussian models}

\label{sec:LGMs}This subsection is devoted to a general class of systems, in
which the TTSHO is a special case. Instead of just two variables, we consider
any number of them ($\xi_{\alpha};~\alpha=1,...,N$), subjected to linearly
restoring deterministic forces and additive Gaussian white noise
($\eta_{\alpha}$). Using a terminology common in climate science, we refer to
these as \textquotedblleft linear Gaussian models\textquotedblright\ (LGMs).
Let us write\footnote{We use the Einstein summation convention for repeated
indices here (super- and sub-script pairs, e.g., $F_{\alpha}^{\beta}\xi
_{\beta}\equiv\sum_{\beta}F_{\alpha}^{\beta}\xi_{\beta}$). In case of the
contrary, the reader will be alerted explicitly. The co- and contra-variant
notation is just for convenience.} the Langevin and Fokker-Planck equations
as, respectively,%
\begin{equation}
\partial_{t}\xi_{\alpha}=-F_{\alpha}^{\beta}\xi_{\beta}+\eta_{\alpha
};~~\left\langle \eta_{\alpha}\left(  t\right)  \eta_{\beta}\left(  t^{\prime
}\right)  \right\rangle =2D_{\alpha\beta}\delta\left(  t-t^{\prime}\right)
\label{LGM-LE}%
\end{equation}
and
\begin{equation}
\partial_{t}P=\partial^{\alpha}\left[  D_{\alpha\beta}\partial^{\beta
}P+F_{\alpha}^{\beta}\xi_{\beta}P\right]  =-\partial^{\alpha}\left(
J_{FP}\right)  _{\alpha} \label{LGM-FPE}%
\end{equation}
Of course, both $D_{\alpha\beta}$ and $F_{\alpha}^{\beta}$ are real and $D$ is
symmetric, while the eigenvalues of both have \textit{positive }real
parts\footnote{In contrast to much of the literature, we write this forces as
$-\mathbb{F}\!\vec{\xi}$ for convenience. With the minus sign, \textit{positivity}
of $\mathbb{F}$ indicates $\vec{\xi}=\vec{0}$ is stable.}. Note that DB is
satisfied if and only if the Fokker-Planck current, $J_{FP}$, vanishes, i.e.,
iff $\left(  D^{-1}\right)  ^{\alpha\gamma}F_{\gamma}^{\beta}$ is symmetric
and positive definite (i.e., $\left(  C^{-1}\right)  ^{\alpha\beta}$ below).
Here, our focus is on general $D$'s and $F$'s which lead to DB violation.

Whether DB is violated or not, the stationary distribution is still a
Gaussian\cite{MLax60,JBW03,ZS07}:%
\begin{equation}
P^{\ast}\propto\exp\left\{  -\frac{1}{2}\xi_{\alpha}\left(  C^{-1}\right)
^{\alpha\beta}\xi_{\beta}\right\}  \label{LGC-P*}%
\end{equation}
Here, $\left(  C^{-1}\right)  ^{\alpha\beta}$ is the inverse of the covariance
matrix $\left\langle \xi_{\alpha}\xi_{\beta}\right\rangle ^{\ast}$ and can be
obtained from $D,F$ either by solving\cite{JBW03} a general version of 
Eqn.(\ref{FC+CF}):%
\[
F_{\beta}^{\gamma}C_{\gamma\alpha}+F_{\alpha}^{\gamma}C_{\gamma\beta
}=2D_{\alpha\beta}%
\]
or summing over the eigenvectors of $F$ \cite{ZS07}. Using the formalism
developed (Eqns.(\ref{DeltaO}-\ref{PAM})), we see that the generalization of
(\ref{L12}),
\begin{equation}
\left\langle \mathcal{L}_{\alpha\beta}\right\rangle ^{\ast}=\left\langle
-\xi_{\alpha}F_{\beta}^{\gamma}\xi_{\gamma}+\xi_{\beta}F_{\alpha}^{\gamma}%
\xi_{\gamma}\right\rangle ^{\ast}=F_{\alpha}^{\gamma}C_{\gamma\beta}-F_{\beta
}^{\gamma}C_{\gamma\alpha}%
\end{equation}
is an anti-symmetric tensor (with $N\left(  N-1\right)  /2$ independent
components). As above, we verify that $\left(  J_{FP}^{\ast}\right)  _{\alpha
}=\frac{1}{2}\left\langle \mathcal{L}_{\alpha\beta}\right\rangle ^{\ast
}\partial^{\beta}P^{\ast}$ so that $\partial_{t}P^{\ast}$ indeed vanishes.

Next, let us turn to a study of the distribution $Q\left(  \mathcal{L}%
\right)  $. Deferring most details to Appendix C, we show only some key
results here (focusing on a single specific pair $\alpha$-$\beta$). First,
its Fourier transform, given by
\[
\tilde{Q}\left(  z\right)  =\int\left[  \exp iz\left(  \xi_{\alpha
}\partial_{t}\xi_{\beta}-\xi_{\beta}\partial_{t}\xi_{\alpha}\right)  \right]
P\left(  \vec{\eta}\right)  P^{\ast}\left(  \vec{\xi}\right)  d\vec{\eta}%
d\vec{\xi}%
\]
can be computed exactly, as the integrand is Gaussian. Using the notation of
$\mathbb{D}$,$\mathbb{F}$, etc., we find $\tilde{Q}=\left[
\det\mathbb{C}^{-1}/\det\mathbb{G}\right]  ^{1/2}$where $\mathbb{G=C}%
^{-1}+2iz\mathbb{XF}+\left(  2z^{2}/\varepsilon\right)  \mathbb{XDX}^{T}$ and
$\mathbb{X}$ is an anti-symmetric matrix with element\footnote{We emphasize
that the subscript $\left(  \alpha\beta\right)  $ should be regarded as a
label which identifies the $\mathcal{L}$ of our focus. The attentive student
will recognize that $\mathbb{X}_{\left(  \alpha\beta\right)  }$ is the
infinitesimal generator of a rotation in the $\alpha$-$\beta$ plane.}
$X_{\left(  \alpha\beta\right)  }^{\mu\nu}=\delta_{\alpha}^{\mu}\delta_{\beta
}^{v}-\delta_{\alpha}^{\nu}\delta_{\beta}^{\mu}$. Thus,
\begin{equation}
\tilde{Q}\left(  z\right)  =\left\{  \det\left[  \mathbb{I}%
+2iz\mathbb{CXF}+\left(  2z^{2}/\varepsilon\right)  \mathbb{CXDX}^{T}\right]
\right\}  ^{-1/2} \label{Q-tilde}%
\end{equation}
But $\det\left[  \mathbb{I}+\mathbb{M}\right]  =1+Tr\mathbb{M}+...+\det
\mathbb{M}$, so that the first two terms of the Taylor series for $\tilde
{Q}^{\ast}$ are%
\[
\tilde{Q}\left(  z\right)  =1-izTr\mathbb{CXF}+...
\]
The first is normalization of $Q\left(  \mathcal{L}\right)  $, while the
second is the first moment, i.e., $-i\partial_{z}\tilde{Q}\left(
0\right)  =\int\mathcal{L}Q$. Verifying $Tr\mathbb{CXF}=-\left\langle
\mathcal{L}_{\alpha\beta}\right\rangle ^{\ast}$, we are confident that
Eqn.(\ref{Q-tilde}) is correct. Proceeding, we find that the $\det$ 
in Eqn.(\ref{Q-tilde}) is a
quartic polynomial (since $\mathbb{X}$ is rank 2), 
\textit{real} in $iz$. The location of its zeros can be
extracted and the inverse transform to $Q$ can be analyzed. The conclusion is
that the decays of $Q$ for $\mathcal{L\rightarrow\pm\infty}$ are dominated by
exponentials:%
\[
\ln Q\left(  \mathcal{L}\right)  \propto-\left\vert \mathcal{L}\right\vert
\]
More crucially, the two slopes are different and accounts for both
$\left\langle \mathcal{L}\right\rangle \neq0$ and $Q\left(  \mathcal{L}%
\right)  -Q\left(  -\mathcal{L}\right)  $ being systematically non-zero.

To end this subsection on LGMs, we present an illustration that contrasts with
the TTSHO in significant ways. In particular, amongst the different ways to
violate DB (by specifying $\mathbb{F}$ and $\mathbb{D}$ so that
$\mathbb{FD\neq DF}^{T}$), the TTSHO involves a real-symmetric $\mathbb{F}$
that does not commute with $\mathbb{D}$. Another class involves
\textit{asymmetric} $\mathbb{F}$'s with complex eigenvalues (with positive
real parts). Without noise, such an $\mathbb{F}$ will bring the system to a
stable focus, i.e., a spiral stable fixed point. With additive white noise,
this condition is \textit{sufficient}, but not necessary, for the stochastic
system to settle into a NESS (details in Appendix D). 
Furthermore, unlike the TTSHO case, it is
possible to reach highly asymmetric\ distributions with this class of LGMs.

\newpage\begin{figure}[pth]
\centering
\includegraphics[width=0.85\textwidth]{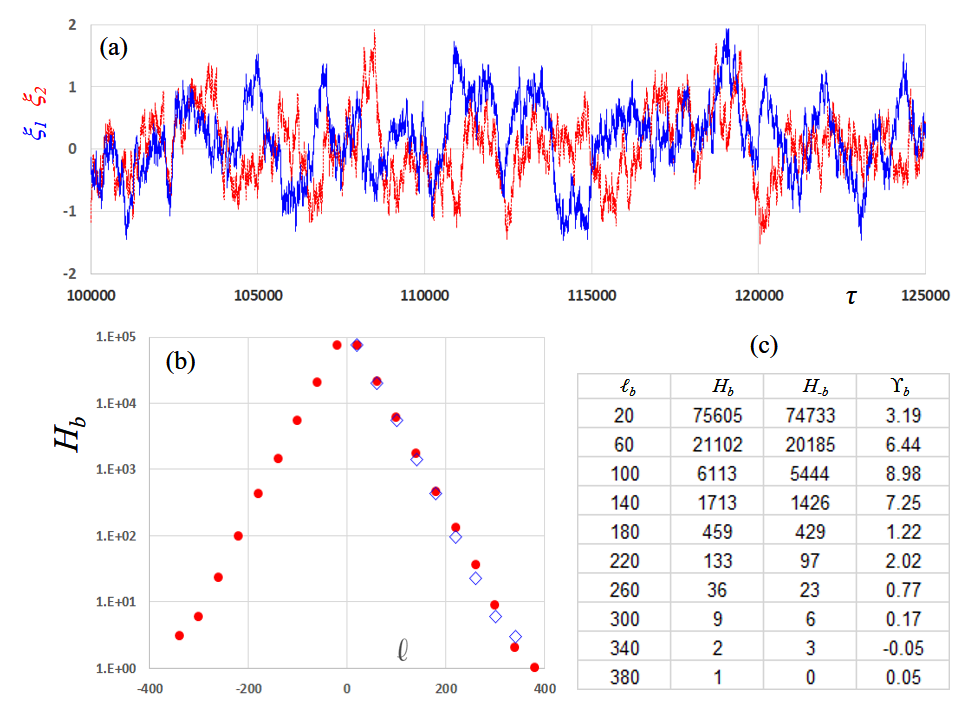}\caption{ Simulation of
an LGM with two degrees of freedom ($\xi_{1,2}$), driven by restoring linear
forces with a spiral and subjected to additive white noise (which models a
single thermal reservoir). (a) A section of the time series, showing no
visually detectable signs of time reversal asymmetry. (b) Histogram of $\ell$
($H_{b}$). Blue diamonds are frequencies for bins with $\ell<0$, showing
systematically lower values. (c) Table of center of bins (value of $\ell$) and
frequencies $H_{\pm b}$, as well as the ratio $\Upsilon_{b}$ (see text for
details). }%
\label{fig:LGM-focus}%
\end{figure}

To illustrate these statements, we present here results from simulations 
of Eqn.(\ref{LGM-LE}), using methods detailed above, in a run of $200K$ 
steps with $\varepsilon=0.001$ for a $N=2$ system. Highlighting the 
non-trivial role played by $\mathbb{F}$ here, we choose%
\[
\mathbb{F}=\left(
\begin{array}
[c]{cc}%
2 & 1\\
-1 & 2
\end{array}
\right)  ,~~\mathbb{D}=\left(
\begin{array}
[c]{cc}%
1 & 0\\
0 & 1
\end{array}
\right)
\]
In other words, this linearly driven system can be interpreted as one being
coupled to a single thermal bath (at some unit $T$). Fig.\ref{fig:LGM-focus}a
shows a section of the time trace of this system in a stationary state, from
which it is again unclear if TRA is present. Similarly, the trajectory in the
$\xi_{1}$-$\xi_{2}$ plane also displays no perceptible rotation around the
origin, as the noise masks the inward spiral due to $\mathbb{F}$. For this
run, $\bar{\ell}$ $\simeq0.935$, which is also quite small compared to
$\sigma_{\ell}\simeq43.9$. However, when we plot the histogram $H_{b}$, as
shown in Fig.\ref{fig:LGM-focus}b, we see that the counts for the $-\ell$ bins
are \textit{systematically} lower than those in the $+\ell$ ones. In
Fig.\ref{fig:LGM-focus}c, we show the bin centers $\ell_{\pm b}$, the
frequencies $H_{\pm b}$, and $\Upsilon_{b}$. Again, we see that every
$\Upsilon_{b}$ is positive and sizable (apart from one with $\left\vert
\Upsilon_{b}\right\vert <1$, associated with a low count). On the theory
front, we verify that $\mathbb{C=I}/2$, so that anti-symmetric part of
$\mathbb{FC}$ gives us $A_{12}=0.5$ and so, $\left\langle \mathcal{L}%
\right\rangle ^{\ast}=1$, a value comparable to the observed $\bar{\ell}$.
Exploiting the techniques detailed in Appendix C, the asymptotic 
slopes of $\ln Q\left(  \mathcal{L}\right)  $ can also be computed and they are
consistent with the decay in $H_{\pm b}$ shown in Fig.\ref{fig:LGM-focus}c.
Thus, we conclude that there is good agreement between simulation data and
theoretical predictions.

\subsubsection{Example in one dimension: a biased random walker with reset}

In this subsection, we focus on a NESS for a system with a one-dimensional
variable, so as to illustrate the TRA signals discussed in subsection
\ref{sec:1D}. As in the previous cases, we will show results for both an
equilibrium system and a NESS, so that some comparisons can be made. The
model here is possibly the simplest version of a random walker with reset 
-- a class of problems which gained considerable attention in recent
years\cite{SatyaReset}. 

Our walker moves on non-negative integer sites along a
line: $i=0,1,...$ in discrete time steps. At each step, it moves either one
site higher (with probability $p$) or `resets' to the origin:%
\[
R\left(  i+1|i\right)  =p;~~R\left(  0|i\right)  =1-p
\]
It is trivial to see that, at large times, the system settles into the
stationary distribution $P^{\ast}\left(  i\right)  =p^{i}\left(  1-p\right)
$. Note that this is identical to the $P^{\ast}$ for a particle in thermal
equilibrium, hopping in a \textquotedblleft gravitational
field\textquotedblright\ with a \textit{impenetrable} floor at $0_{-}$. All we
need to arrive at that equilibrium state is to impose DB satisfying rates,
i.e., the ratio for hopping \textquotedblleft up one rung\textquotedblright\ to
hopping \textquotedblleft down a rung\textquotedblright, 
$R\left(  i\rightarrow i+1\right)  /R\left( i\rightarrow i-1\right) $, to be $p$.

Meanwhile, DB violation in our NESS is clear, as the persistent net
currents are%
\[
K^{\ast}\left(  i+1|i\right)  =p^{i+1}/\left(  1-p\right)  ;~~K^{\ast}\left(
0|i\right)  =p^{i}%
\]
and satisfy $K^{\ast}\left(  i+1|i\right)  =K^{\ast}\left(  0|i\right)
+K^{\ast}\left(  i|i-1\right)  $ for $i>0$ and $K^{\ast}\left(  1|0\right)
=\sum_{i>0}K^{\ast}\left(  0|i\right)  $. There are many non-trivial current
loops, of course, all of the form $0\rightarrow1\rightarrow...\rightarrow
m\rightarrow0$. with $m>1$. The TRA generated can be detected by studying
$\tilde{\Gamma}$ introduced in Eqn.(\ref{Gamma-tilde}). Here, $\zeta$ is the
set of non-negative integers, $i$, and the simplest illustration for
$\tilde{\Gamma}$ is the single time step case: $\upsilon=1$. Then, $P\left(
j,1|i,0\right)  P^{\ast}\left(  i\right)  $ reduces to
\begin{align*}
R\left(  j|i\right)  P^{\ast}\left(  i\right)   &  
=\left[  p\delta (j,i+1)+\left(  1-p\right)  \delta (j,0) \right] P^{\ast}\left( i\right) \\
&  =\left[  p\delta (j,i+1)+\left(  1-p\right) \delta (j,0)\right] p^{i}\left(  1-p\right)
\end{align*}
while the integral reduces to the sum%
\[
\sum_{i,j}ji\left(  j-i\right)  P\left(  j,1|i,0\right)  P^{\ast}\left(
i\right)  =\sum_{i=1}\left(  i+1\right)  ip^{i+1}\left(  1-p\right)
\]
The result is readily obtained:%
\[
\tilde{\Gamma}\left(  1\right)  =2\left(  \frac{p}{1-p}\right)  ^{2}%
\]
Meanwhile, the standard deviation is $\sigma_{\zeta}=\sqrt{p}/\left(
1-p\right)  $, so that the dimensionless measure is%
\[
\tilde{\Gamma}/\sigma_{\zeta}^{3}=2\sqrt{p}\left(  1-p\right)
\]
the maximum of which occurs at $p=1/3$.

\begin{figure}[pth]
\centering
\includegraphics[width=0.85\textwidth]{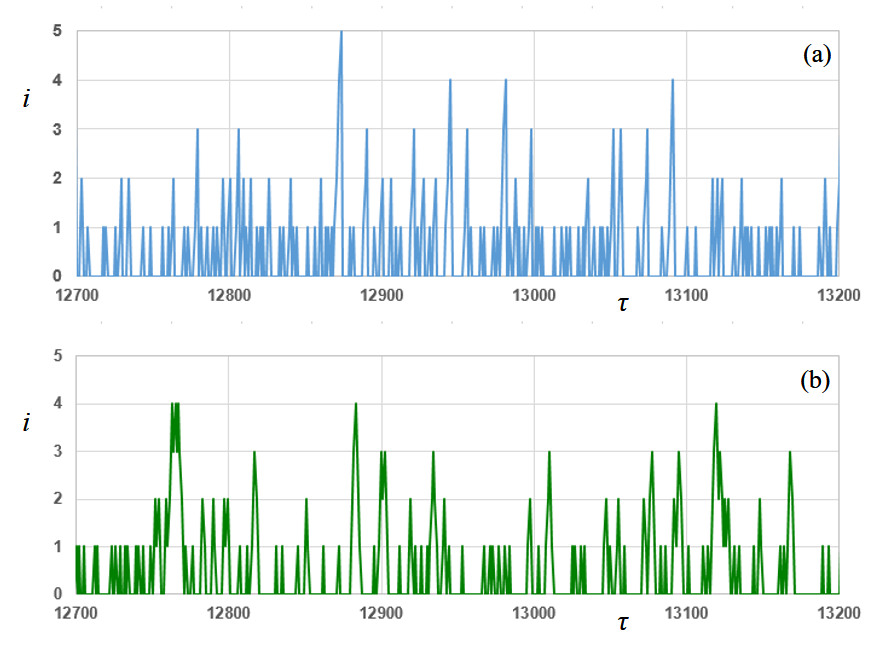}\caption{ The time series for
(a) a random walker with reset ($p=1/3$) in the steady state. Over the $500$
steps shown, it is discernibly asymmetric under time reversal. See text for
the values of the mean position, its variance, the asymmetry $\tilde{\Gamma}$,
and the ratio $\tilde{\Gamma}/\sigma_{\zeta}^{3}$ (over this run of $50K$
steps). By contrast, (b) shows a walker stepping up and down with probability
${\frac{1}{4}}$ and ${\frac{3}{4}}$, respectively, in a time symmetric
equilibrium state. Note that both walkers have the \textit{same} 
$P^{\ast}$, as hinted by these traces. }%
\label{fig:1D}%
\end{figure}

Simple simulations of this process provides excellent agreement with these
predictions. For example, Fig.\ref{fig:1D}a illustrates a part of the time
trace, $i\left(  \tau\right)  $, in a run with just $50K$ steps with $p=1/3$.
For this run, the comparison between data and theoretical predictions for 
various quantities (mean $\bar{\zeta}$, SD $\sigma_{\zeta}$, asymmetry 
$\tilde{\Gamma}$, and ratio $\tilde{\Gamma}/\sigma_{\zeta}^{3}$) are shown:

\begin{center}
\begin{tabular}{ | m{2.2cm} | m{1.5cm} | m{1.5cm} | m{1.5cm}| m{1.5cm} | } 
  \hline
 & $~~~~\bar{\zeta}$ &~~~ $\sigma_{\zeta}$ & $~~~~\tilde{\Gamma}$ &~~ $\tilde
{\Gamma}/\sigma_{\zeta}^{3}$ \\ 
  \hline
Simulations &~ $0.5002$ &~ $0.8658$ &~ $0.4996$ &~ $0.7698$ \\ 
  \hline
~~~Theory &~ $0.5000$ &~ $0.8660$ &~ $0.5000$ &~ $0.7698$ \\ 
  \hline
\end{tabular}
\end{center}
So we conclude that the agreement is excellent and TRA is clear. 
As a contrast, in a similar run with DB satisfying rates 
(Fig.\ref{fig:1D}b, with specifically 
$R\left( i\rightarrow i+1\right)  =p/\left(  1+p\right)  $ and  
$R\left( i\rightarrow i-1\right)  =1/\left(  1+p\right) $ ), the asymmetry is 
found to be zero, while the average and variance are entirely consistent with 
theoretical values. To emphasize, the stationary $P^{\ast}\left(  i\right) $
for both systems are identical. There are non-trivial current loops in the
NESS, as measured by $\tilde{\Gamma}$, but none in the equilibrium system.
Visually, it is possible to discern the difference (in TRA) between these two
time traces. Of course, the NESS in this example is quite extreme. If a
combination of these two dynamics were introduced (i.e., with a fraction of
the updates being DB violating), then it may be quite difficult to distinguish a
system \textquotedblleft slightly perturbed from equilibrium\textquotedblright%
\ than one truly in equilibrium.

\subsection{From the subtle to the manifest}

All of the examples studied above are solvable \textit{exactly}. In this
sense, they may be regarded as \textquotedblleft toy models\textquotedblright%
\ rather than realistic ones designed for systems in nature, where
non-linearities abound. In general, the latter are not analytically accessible,
so that information about how they display TRA is typically gleaned from
simulations. Of course, we can also study data collected from physical
systems\textit{ directly}, such as those under controlled environments in
laboratories or observed in complex natural settings (from micro to global, in
e.g., biological and climate sciences\cite{BBF16,WFMNZ21}). In nearly all such
systems, it is the exception that the underlying dynamics obey DB and the
steady states fall within equilibrium statistical mechanics. While some of
these NESS are \textquotedblleft close\textquotedblright\ to equilibrium and
display subtle signals of TRA, others are clearly \textquotedblleft far
from\textquotedblright\ equilibrium and TRA is manifest. In this subsection,
we illustrate this wide spectrum of possibilities with examples from both
physical data and numerical simulations.

\subsubsection{Illustrations from two climate systems}

Let us begin with two illustrations of TRA published recently\cite{WFMNZ21},
based on data from our climate. Both display somewhat \textquotedblleft
subtle\textquotedblright\ asymmetry under time reversal, in that one shows
$\bar{\ell}<\sigma_{\ell}$ while the other, $\bar{\ell}\simeq\sigma_{\ell}$.
The former is associated with the well-known phenomenon of El-Ni\~{n}o and the
latter, with the Madden-Julien Oscillation. Both are considerably further from
equilibrium than the simple examples shown above (where $\bar{\ell}\ll
\sigma_{\ell}$).

\begin{figure}[pth]
\centering
\includegraphics[width=0.85\textwidth]{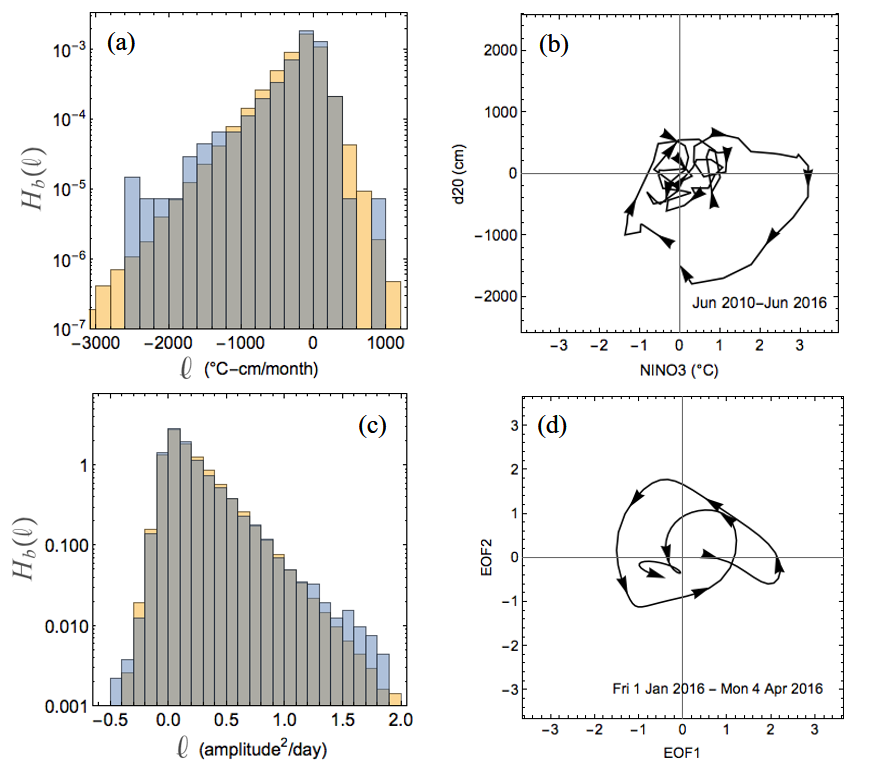}\caption{
Histograms/distributions of $\ell$ and trajectories for a pair of quantities
associated with El-Ni\~{n}o (a,b) and Madden-Julien Oscillations (c,d). 
In (a,c), blue bars represent data, orange ones represent theory, 
and grey is for regions of overlapping blue/orange. See text for details. 
(Figures reproduced from Ref.\cite{WFMNZ21}.) }%
\label{fig:Climate}%
\end{figure}

El-Ni\~{n}o is generally associated with the warming of the tropical Pacific
Ocean. A less well-known feature is the variations of the depth of the
thermocline\footnote{Thermocline is a region below the ocean surface where
temperatures change rapidly. See, e.g.,
https://en.wikipedia.org/wiki/Thermocline.}. While there are many ways to
characterize \textquotedblleft warming\textquotedblright\ and
\textquotedblleft depth,\textquotedblright\ we focus on two common measures of
these features:

\begin{itemize}
\item NINO3, the sea surface temperature averaged over a certain region in the
eastern Pacific (in units of $^{\circ }C$)

\item d20, the depth of the $20^{\circ }C$ isotherm in the tropical Pacific
(in units of $cm$).
\end{itemize}
Details of these measures and the data sets chosen may be found in Ref.
\cite{WFMNZ21}. The time series of these quantities consist of monthly
averages of observations from 1960 to 2016. Setting the means of these series
to zero, we work with the \textquotedblleft anomalies\textquotedblright\ of
NINO3 and d20. When the\ entire trajectory is plotted in the NINO3-d20 plane,
there is no obvious systematic rotation around the origin. In 
Fig.\ref{fig:Climate}b, we illustrate with a small section, which may give an
impression of a tendency to rotate clockwise. When we compute $\bar{\ell}$ and
$\sigma_{\ell}$, we find $-205.0$ and $432.4$ (in units of $^{\circ }%
C$-$cm/month$), respectively. Since $\bar{\ell}$ is just half of $\sigma
_{\ell}$, we compile the histogram of $\ell$'s and display it (colored as blue
columns) in Fig.\ref{fig:Climate}a. There is a clear asymmetry in favor of
negative $\bar{\ell}$ bins. Similarly, in Figs. \ref{fig:Climate}c and d, we
show the distribution and a sample trajectory for the two dominant amplitudes
associated with the Madden-Julien Oscillation. Here, the asymmetry is much
more prominent in both representations, a result consistent with $\bar{\ell}$
being comparable to $\sigma_{\ell}$, namely, $0.201$ and $0.263$, respectively
(in units of $amplitude^{2}/day$).

On the theory front, parameters for LGMs were chosen to fit the time-series.
(See Appendix E for further details.) In turn, these models predict values for
$\bar{\ell}$ and $\sigma_{\ell}$ and, correponding to the above cases, they 
are $-205.7$, $350.7$, $0.199$, and $0.250$. All predictions are in acceptable 
agreement with data. The distributions $Q$ can be computed by inverting 
$\tilde{Q}$ numerically. The resultant are plotted as orange columns in 
the histogram figures ($H_{b}$). Meanwhile, the grey regions in the columns 
represent the overlap of theory and data. From these, we conclude that 
the LGMs have captured the essentials of the asymmetry in $H_{b}$ 
(and $Q\left(\mathcal{L}\right)  $), and they provide good approximants for the
non-equilibrium quasi-stationary states in nature. Finally, let us highlight
the fact that the LGMs for both systems belong to the class of $\mathbb{F}$'s
with stable focus. Unlike the model system presented in Section
\ref{sec:LGMs}, the twists (i.e., antisymmetric parts of $\mathbb{F}$) are
progressively stronger (from El-Ni\~{n}o to MJO). The underlying physics here
is entirely consistent with the increase in both the ratio $\bar{\ell}%
/\sigma_{\ell}$ and the prominence of the asymmetry in the $H_{b}$ plots.

\subsubsection{Hopf bifurcation and transition from subtle to manifest display
of TRA}

\label{sec:Hopf}While the climate data we presented represent a significant
increase of the TRA signal from that in the TTSHO, say, we may still regard
them as borderline cases in the \textquotedblleft
subtle-manifest\textquotedblright\ spectrum. The next level of asymmetry is so
prominent that it deserves the term \textquotedblleft
manifest,\textquotedblright\ namely, when the system undergoes a Hopf
bifurcation\cite{Hopf43} and displays limit cycles which are
unmistakably irreversible in time. In nature, such phenomena are abundant --
in \textit{quasi-stationary} cyclic states rather than truly stationary ones
-- from predator-prey systems to chemical reactions. To explore bona-fide
stationary stochastics processes, we turn to model systems inspired by various
natural phenomena. In this subsection, we follow this route, in order to
illustrate the effects of such a transition on $\left\langle \mathcal{L}%
\right\rangle $ and its distribution. Naturally, we must go beyond the LGM and
so, typically, few exact analytic results are known and simulations offer the
best progress for most model systems. By exploiting a sufficiently simple
model\footnote{Perhaps the \textit{simplest} model available, the
deterministic part of this system is known as the \textit{normal form} of a
Hopf bifurcation.} for a stochastic (supercritical) Hopf bifurcation, we can
present some exact results (stationary distribution $P^{\ast}$ and current
$\vec{J}_{FP}^{\ast}=\vec{\mu}P^{\ast}-D\vec{\nabla}P^{\ast}$) as well as some
visually appealing simulation data.

Here again, our model consists of two variables ($\xi_{1,2}$), forced by a
conservative part ($-\vec{\nabla}V$) plus a twist ($\vec{F}_{\omega}%
\propto\omega$), and subjected to a trivial $\vec{\eta}$ ($\mathbb{D}%
=T\mathbb{I}$). Defining $r^{2}\equiv\xi_{1}^{2}+\xi_{2}^{2}$, we choose a
standard potential%
\[
V\left(  \xi_{1},\xi_{2}\right)  =\frac{\rho}{2}r^{2}+\frac{u}{4}r^{4}%
\]
which allows a Hopf bifurcation, at $\rho=0$. Without the twist, the system
settles into thermal equilibrium of course (with $P^{\ast}\propto e^{-V/T};$
$k_{B}=1$) and is often used to model\footnote{The Landau model.}, say, a
system of classical spins. For $\rho>0$, it is \textquotedblleft above the
critical temperature\textquotedblright\ where the rotational symmetry is
unbroken. When $\rho$ turns negative, we have spontaneous symmetry breaking,
with the minimum energy state given by $r =\sqrt{-\rho/u}$. If a twist
\[
\vec{F}_{\omega}=\left(
\begin{array}
[c]{cc}%
0 & \omega\\
-\omega & 0
\end{array}
\right)  \vec{\xi}%
\]
is added, then the stochastic system, specified by the Langevin equation,
\[
\partial_{t}\vec{\xi}=-\vec{\nabla}V-\vec{F}_{\omega}+\vec{\eta}%
\]
will tend to \textquotedblleft jiggle\textquotedblright\ near the origin
$\vec{\xi}=\vec{0}$ in case $\rho>0$ (similar to the example shown in 
Fig.\ref{fig:LGM-focus}). But, for $\rho<0$ (with large $-\rho/u$ and small $T$),
the trajectories will be mostly circular with small deviations. We may refer
to such cases as limit cycles with low noise. Near the transition, i.e., small
$-\rho/u,\omega$ and moderate $T$, TRA signals from the system will 
undoubtedly be \textquotedblleft subtle\textquotedblright . Deep into 
the broken-symmetry and twist-driven region, i.e., large 
$-\rho/u,\omega$ and small $T$, the signal should be 
\textquotedblleft manifest\textquotedblright .
In this toy model, it is straightforward to show that, 
for the Fokker-Planck equation%
\[
\partial_{t}P=-\vec{\nabla}\cdot\vec{J}_{FP}=\vec{\nabla}\cdot\left[  \left(
\vec{\nabla}V+\vec{F}_{\omega}\right)  P+T\vec{\nabla}P\right]
\]
the stationary distribution is simply $P^{\ast}\propto e^{-V/T}$
(since both $\vec{\nabla}\cdot\vec{F}_{\omega}$ and $\vec{F}_{\omega}\cdot
\vec{\nabla}V$ vanish), with $\vec{J}_{FP}^{\ast}=-\vec{F}_{\omega}P^{\ast}$.
From here, quantities such as $\left\langle \mathcal{L}\right\rangle ^{\ast
},\sigma_{\mathcal{L}},Q^{\ast}\left(  \mathcal{L}\right)  $ can be computed
numerically, as none of the integrands are simple Gaussians.

Our main interest here is to illustrate the qualitative features of this
stochastic process and how they differ from those in the LGMs. Thus, we will
present only a figure of a simulation of the above system and highlight the
differences, while avoiding technical details. As in previous simulations,
we carried out a run of $200K$ steps with $\varepsilon=0.001$ with
$\rho=u=\omega=1$ and $T=0.1$.

\begin{figure}[pth]
\centering
\includegraphics[width=0.85\textwidth]{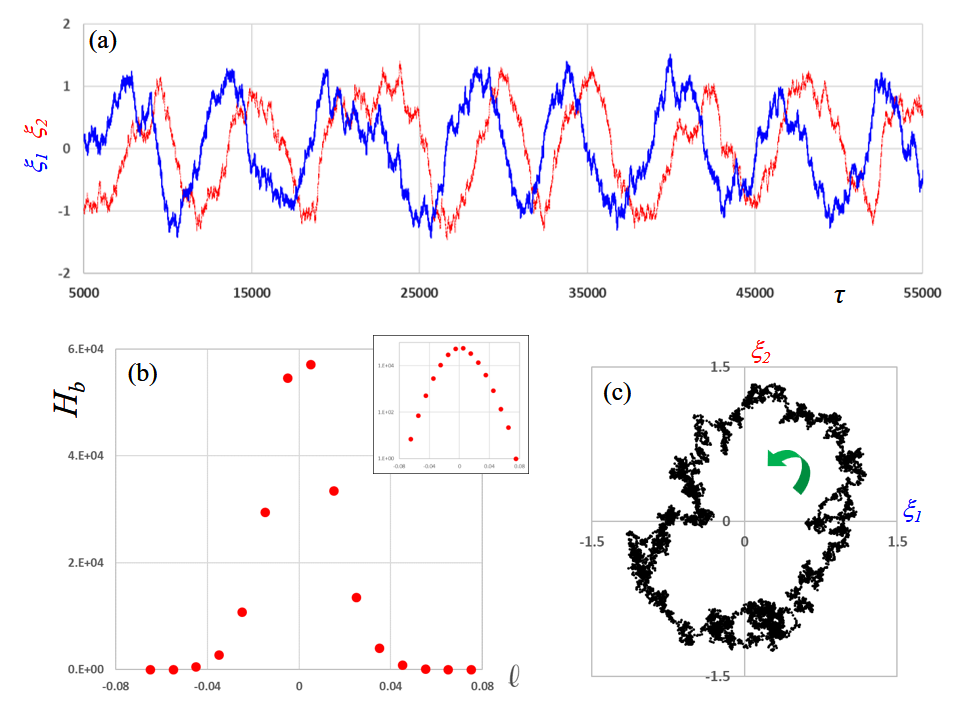}\caption{ Simulation of a system
with two degrees of freedom ($\xi_{1,2}$), driven into a limit cycle by
deterministic forces and subjected to additive white noise (which models a
single thermal reservoir). (a) A section of the time series, displaying
prominent signs of time reversal asymmetry. (b) Histogram of $\ell$ ($H_{b}$).
The asymmetry between positive and negative bins is quite clear. In the inset
is a log plot, showing the distribution is closer to 
Gaussian rather than (double) exponential. 
(c) A section of the trajectory in the $\xi_{1}$-$\xi_{2}$ plane,
with the green arrow indicating the direction of motion. }%
\label{fig:LC}%
\end{figure}

Fig. \ref{fig:LC}a shows a section of the time trace of this system in a
stationary state, from which it is quite clear that TRA is present. For
example, it is qualitatively similar to a predator-prey system, in which a
rise of the predator population ($\xi_{2}$, red trace) soon follows a rise in
the prey numbers ($\xi_{1}$, blue trace). Similarly, the trajectory in the
$\xi_{1}$-$\xi_{2}$ plane displays an obvious direction of motion, 
as indicated by the green arrow. 
Despite the clear sense of rotation in this run, the presence of
angular momentum is not prominent, as its typical values are dominated by the
noise in $\partial_{t}\vec{\xi}$. As a result, $\bar{\ell}$ $\simeq1.00$ is
relatively small compared to $\sigma_{\ell}\simeq14.3$. While the histogram
$H_{b}$ in Fig.\ref{fig:LC}b is certainly biased in favor of $\ell>0$, the
asymmetry is overwhelmed by the width of the distribution. More interesting is
the inset figure, showing that $H_{b}$ is no longer exponential at large
$\ell$. Instead, it is essentially Gaussian, so that its make-up can be
interpreted readily: a noise dominated Gaussian part, displaced by a small
positive mean ($\bar{\ell}>0$). Not surprisingly, when $T$ is lowered by a
factor of $100$, say, $\sigma_{\ell}$ drops by a factor of $10$, while
$\bar{\ell}$ remains essentially unchanged. The stationary distribution
($P^{\ast}$, not displayed explicitly) is quite distinct from the LGM case
above. Instead of a single peaked Gaussian, it resembles a volcano, with a dip
at the center and a rim (circular in this simple model) at $r=1$. 
There is a non-trivial, persistent probability
current moving around this structure, contributing to a non-vanishing
$\left\langle \mathcal{L}\right\rangle ^{\ast}$. It is reasonable to conclude
that, when a stochastic process settles into a (noisy) limit cycle, DB and
time reversal violation will be \textquotedblleft manifest.\textquotedblright

\subsection{Astonishing complex behavior from minimal models in nonequilibrium
steady states}

\label{sec:DWRLG}The simple solvable examples above should not give the reader
the impression that NESS is well understood. On the contrary, a casual glance
around us provides an astounding array of phenomena which we cannot predict
from their microscopic constituents (and interactions), e.g., all forms of
life. Even much simpler systems, such as point particles diffusing on a
lattice under DB violating rules, can pose serious challenges. Known as driven
diffusive systems, they have been the focus of research since the
1980's\cite{SZ95}. Arguably the simplest of these is the well-established
Ising lattice gas\cite{YL52} driven far from equilibrium\cite{KLS84}. It
displays many intriguing properties\footnote{These include DB violation of
course. But TRA is so \textquotedblleft obvious\textquotedblright\ that it was
never quantitatively studied in detail.}, some of which are yet to be
understood. Here, we present another \textquotedblleft minimal
model\textquotedblright\ -- the lattice gas version\cite{DS95} of the
Widom-Rowlinson model\cite{WR70} -- easily understood if subjected to
equilibrium conditions. However, when driven uniformly out of equilibrium
(e.g., biased diffusion in one direction, say), the system displays remarkably
surprising and complex behavior\cite{DZ18,LDZ21}. In this subsection, we
present only a brief summary, so as to illustrate how \textit{seemingly
trivial }changes of the dynamics (from DB satisfying to violating) can lead to
profound changes in the collective behavior of a statistical system.

\begin{figure}[pth]
\centering
\includegraphics[width=0.85\textwidth]{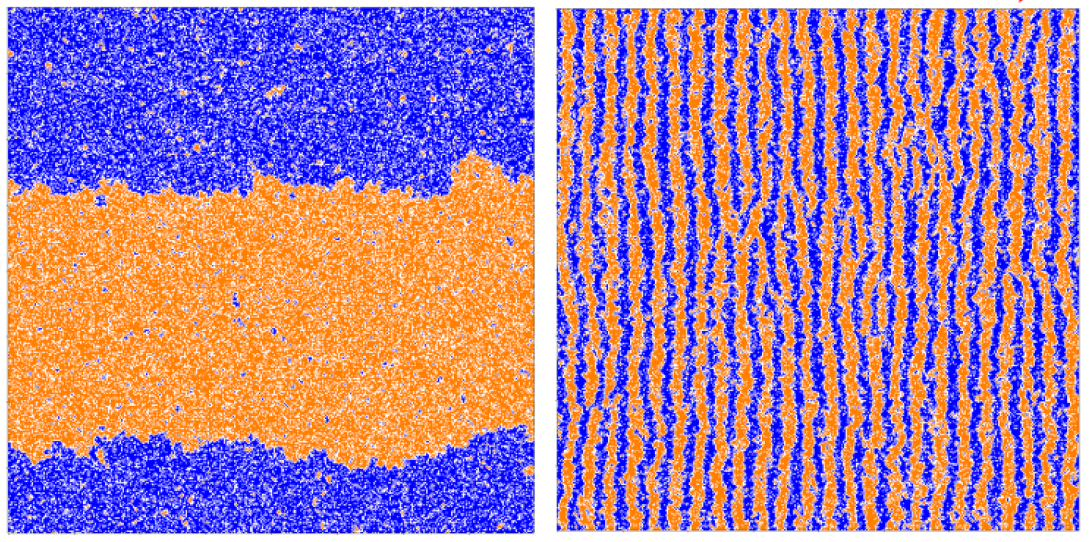}\caption{ Typical
configurations for the Widom-Rowlinson lattice gas (left panel) and the driven
version (right panel, with particles biased $3\times$ more to move to the right
than to the left). Blue and orange sites are occupied by equal numbers of 
the two species of particles. Vacancies are shown as white. 
Both systems are $400 \times400$ with 
$\bar{\rho}=0.7$. (Figures reproduced from Ref.\cite{LDZ21}.) }%
\label{fig:DWRLG}%
\end{figure}

Introduced in 1995, the Widom-Rowlinson lattice gas (WRLG) \cite{DS95} 
consists of two species of particle
(say, A and B) placed singly on a square lattice (or other lattices, in other
dimensions), diffusing freely via nearest neighbor particle-hole exchange. 
Configurations are specified by 
$\left\{  s\left(  \vec{x}\right)  \right\}  ;s=\pm1,0$
representing site $\vec{x}$ being occupied by A,B, or vacant. The only other
constraint is that AB nearest neighbor pairs are excluded. With \textit{no}
energy functional, temperature is \textit{irrelevant}. The only control
parameters are the densities of the two species. In other words, entropy is
the sole key for this equilibrium system (i.e., the microcanonical ensemble
with $P^{\ast}\left(  \mathcal{C}\right)  \propto1$ for any allowed
$\mathcal{C}$). Since the NN-exclusion acts as an entropic force, we may
expect an effective attraction between AA and BB pairs, as in a standard Ising
lattice gas. This picture is mainly confirmed in simulation studies, most of
which have equal numbers of A and B, with the overall density ($\bar{\rho}$)
as a tunable parameter. As expected, the system is homogeneous for small
$\bar{\rho}$, with a transition to a phase separated state (two regions:
A-rich or B-rich, e.g., left panel in Fig.\ref{fig:DWRLG}) when the density
exceeds a critical value: $\rho_{c} \simeq0.617(1)$. Near $\rho_{c}$ the
critical behavior is in the Ising universality class\cite{DS95}. Our interest
here is when a bias is introduced into the particle-hole exchange so that DB
is violated and the system settles into a NESS instead.

Unlike earlier studies of biased diffusion of two species, in which
interesting phases were discovered when A and B are driven in
\textit{opposite} direction\cite{ABC}, \textit{both} species are driven in the
\textit{same} direction in this system\footnote{For example, in the maximal
drive case, a NN or NNN particle-hole pair is allowed to exchange only when
$x_{hole}=x_{particle}+1$ (provided the constraint is satisfied),
\textit{regardless} of whether the particle is $A$ or $B$.}. Intuitively, it
appears that a drive that does not distinguish the species should have little
effect. Yet, completely unexpected properties were discovered through
simulations, even when the system is in the \textit{disordered} phase. We
summarize some of these findings here.

If the single species, non-interacting Ising lattice gas
is driven, the stationary distribution remains uninteresting ($P^{\ast}%
\propto1$).\footnote{This follows from \textquotedblleft pairwise detailed
balance.\textquotedblright\ Though $R\left(  i|j\right)  \neq R\left(
j|i\right)  $ in general, we can identify unique pairs of transitions with
equal rates: $R\left(  i|j\right)  =R\left(  k|i\right)  $. Thus, $P^{\ast}=1$
satisfies $\Delta_{t}P^{\ast}=0$.} For the driven WRLG in \textit{one
}dimension, it is clear that the same mechanism leads to $P^{\ast}\propto1$.
However, simulations in \textit{two} dimensions (e.g., on square lattices of
$L\times L$ sites, with toroidal BCs) show that $P^{\ast}$ is far from 
simple. Generically, in addition to displaying long range correlations 
(as in the KLS model\cite{KLS84}), the drive induces a 
\textit{preferred length}\cite{DZ18,LDZ21}.
In particular, even for $\bar{\rho}\ll 1$, the
steady state structure factor (i.e., the Fourier transform of the two point
equal-time correlation ($\left\langle s\left(  \vec{x}\right)  s\left(  \vec{x}%
^{\prime}\right)  \right\rangle ^{\ast}$) in this homogeneous state does not
assume the standard Ornstein-Zernike form. Instead, it \textit{peaks} at a
preferred, non-trivial wave-vector, $\vec{q}$, \textit{parallel} to the drive
direction. Correspondingly, $\left\langle s\left(  \vec{x}\right)  s\left(
\vec{x}^{\prime}\right)  \right\rangle ^{\ast}$ does not decay 
exponentially\footnote{Instead, with a power law envelope.}
for large $\left\vert \vec{x}-\vec{x}^{\prime}\right\vert $, but
oscillates much like a damped spring, with wavelength
$\Lambda\equiv2\pi/\left\vert \vec{q}\right\vert $. For $\bar{\rho}=0.5$,
$\left\vert \vec{q}\right\vert $ rises more or less linearly with the strength
of the drive\cite{LDZ21}. For fixed drive and increasing $\bar{\rho}$,
$\vec{q}$ also increases, but only slightly. Meanwhile, the value at the peak
rises substantially with higher $\bar{\rho}$, until a critical $\bar{\rho}%
_{c}$ is reached. Beyond $\bar{\rho}_{c}$, the peak height scales with system
size rather than being intensive -- a characteristic of long range order
and phase separation. Unlike the undriven case, where the system always
ordered into two particle-rich regions (similar to the Ising case, i.e.,
$\Lambda=L$), the driven WRLG displays lamellae \textit{normal to the drive},
with a \textit{fixed} width -- an $L$-independent $\Lambda/2$. 
To illustrate, we show two typical configurations with $\rho=0.7$ in 
Fig.\ref{fig:DWRLG}, one in equilibrium and the other, a driven system.

Further, $\Lambda$ varies in a very perplexing manner. To emphasize, this
lamellae structure is present even in the homogeneous phase with a small
drive (though difficult to discern). 
For maximal drive, $10\lesssim\Lambda\lesssim12$ (lattice spacings,
deduced from the correlation function) deep within this phase ($\bar{\rho
}\sim0.4$). Around $\bar{\rho}_{c}\sim0.77$, ordering emerges and the lamellae
appear as \textquotedblleft visible strips\textquotedblright\ (in square
lattices). As $\bar{\rho}$ increases further, $\Lambda$ rises by minute
amounts and remains $L$ independent. In other words, in these regime,
$\Lambda$ appears to be a characteristic of the microscopic parameters, so
that the number of strips is essentially $L/\Lambda$. As $\bar{\rho}_{c}$
closes on unity, this number decreases rapidly, via a series of
\textquotedblleft strip-merging\textquotedblright\ transitions. The boundary
of this regime, $\bar{\rho}_{sm}$, is likely to depend on $L$ in a complicated
manner. Eventually ($\bar{\rho}\lesssim 1-2/L$, clearly), the system ends in a
completely phase segregated state: $\Lambda=L$, with one strip of each
species, separated by thin ($\sim$ lattice spacing) strips of holes. In
simulations involving $L\leq400$, strip-merging appears for $\bar{\rho}%
>\bar{\rho}_{sm}\sim0.9$. For other values of the bias, the behavior of
$\Lambda,\bar{\rho}_{c},\bar{\rho}_{sm}$ on $\bar{\rho}$ is highly non-trivial
\cite{DZ18,LDZ21} and remains to be explored in detail. In all cases,
$\Lambda$ rises to $O\left(  L\right)  $ at both extremes of $\bar{\rho}$.

The driven WRLG provides a good example of a stationary state which is clearly
far from having the maximal entropy (disorder). No one could have predicted
the emergence of the lamellae structures from the simple transition rates
(translation invariant in both space and time) in this minimal system. 
Although we did not discuss TRA explicitly in this case, the drift of particles
is so clear that showing TRA by some other means seems redundant. What needs
to be emphasized here is the DB violation is undeniably at play, as the steady 
state display phenomena which are far from being consistent with maximal entropy. 
In many ways, the surprises here are in the same class as those found in, say,
Conway's game of life\cite{Conway}, in which complex patterns emerge from
minimal rules and constituents. We believe that, beyond simple signals of time
reversal asymmetry, this aspect of non-equilibrium statistical mechanics --
the emergence of complexity from simplicity -- will be a rewarding and
exciting frontier for future investigations into the mysteries of 
cooperative phenomena.

\section{Summary and Outlook}
\label{sec:SO}

When a statistical system settles into its stationary state, time translational
invariance holds, by definition. However, this state may or may not be 
symmetric under time \textit{reversal}, depending on whether the system is
in thermal equilibrium or a non-equilibrium steady state. In layman's terms,
snap shots of a movie at different times -- of either steady states -- will 
appear statistically the same, 
but there may be a difference when the movies are run forwards
\textit{vs.} backwards. The origin of the difference lies in whether the rules
for evolution of the system obeys detailed balance or not. This article is
devoted to simple ways to detect asymmetries under time reversal.

In general, to observe time dependence phenomena, we need correlations at
\textit{unequal} times. Here, we presented the simplest of such
functions to detect TRA, relying on the most elementary of observables and the
minimal separation of just \textit{two} times. In elementary physics, these
observables can be positions of particles ($\vec{x}$) while \textquotedblleft
minimal time separation\textquotedblright\ can be $\delta t$. Meanwhile, the
velocity $\vec{v}=\delta\vec{x}/\delta t$ is the simplest quantity odd under
time reversal. To study the statistical properties of systems in
\textit{steady states}, we turn to averages (means and correlations), 
denoted by  $\left\langle ...\right\rangle ^{\ast}$. 
But\footnote{Restricting ourselves to finite systems, we can always choose
its origin so that $\left\langle \vec{x}\right\rangle ^{\ast}$ vanishes.
Meanwhile, it is rare for steady states to have 
$\left\langle \vec{v}\right\rangle ^{\ast}\ne 0$. } 
$\left\langle \vec{x}\right\rangle ^{\ast} = \left\langle \vec{v}\right\rangle ^{\ast}\equiv0$, 
and to detect time reversal asymmetry, nothing
is simpler than the averages of their product. Of
course, $\left\langle \vec{x}\cdot\vec{v}\right\rangle ^{\ast}=$ $\frac{1}%
{2}\left\langle \partial_{t}x^{2}\right\rangle ^{\ast}\equiv0$, and we are
naturally led to consider $\left\langle \vec{x}\times\vec{v}\right\rangle
^{\ast}$ (or $\left\langle \xi_{\alpha}\Delta\xi_{\beta}-\xi_{\beta}\Delta
\xi_{\alpha}\right\rangle ^{\ast}$ in general). Recognizing $\vec{x}\times
\vec{p}$ is angular momentum in elementary physics, we coin the term
\textit{probability} angular momentum ($\mathcal{L}$) for the correlation
function introduced. If configuration space consists of just one dimension,
then $\vec{x}\times\vec{p}\equiv0$ and a different quantity (involving third
moments) was proposed for detecting TRA.

Relating these abstract concepts to data (from simulations, experiments, or
observations), we are aware that averages from finite data sets are almost
never identically zero, so that using $\left\langle \mathcal{L}\right\rangle
^{\ast}\neq0$ as a test for TRA is too idealistic. Further, we find that
comparing its magnitude to its standard deviation ($\sigma_{\mathcal{L}}$)
rarely provides a reliable gauge. In particular, many model systems
with built in DB violating dynamics display $\left\langle \mathcal{L}%
\right\rangle ^{\ast}\ll\sigma_{\mathcal{L}}$. This issue motivated us to
consider $Q\left(  \mathcal{L}\right)  $, the full distribution of
$\mathcal{L}$. For many system, it is broadly spread around $\mathcal{L}=0$, 
while TRA manifests as an asymmetry: 
$Q\left(  \mathcal{L}\right)  \ne Q\left( -\mathcal{L}\right)  $. 
In practice, time series of observables can be used to
build histograms associated with $\mathcal{L}$, while we proposed a measure
($\Upsilon$) which appears to be quite reliable for deciding if a system
displays TRA or not. We refer to systems with such signals of TRA as
\textquotedblleft subtle,\textquotedblright\ since there are no overt signs of
time reversal imbalance. By contrast, many steady state systems in nature (or
quasi-stationary states) display \textquotedblleft manifest\textquotedblright%
\ signs of TRA, especially those with limit cycles. At the far end of this
subtle-manifest continuum are stochastic processes which display phenomena
well beyond our knowledge or expectations: Discoveries of new
species of life forms continue to astonish us. Even seemingly trivial
constituents evolving with simple rules can lead to unpredictably complex
behavior, e.g., Conway's game of life. Driven diffusive systems\cite{SZ95} provide
another good example, as we learn that collective phenomena in stochastic
processes evolving with DB violating dynamics can be extremely rich and
surprising. To illustrate these ideas, a number of examples were presented,
from simple solvable systems to complex physical ones.

The limited study here should be regarded as a small glimpse into the wide
vistas of NESS. Specifically, our explorations naturally raised more 
questions. In addition to the ones noted in the text, we list a few 
other examples here. (i) The simplest of observables 
for detecting TRA is considered here, as $\mathcal{L}$ is a two point 
correlation at \textquotedblleft one time step\textquotedblright\ apart, 
though $Q\left( \mathcal{L}\right)  $ effectively involve all its higher 
moments. One generalization we did not study is correlations involving larger
separations of times,\ e.g., $\mathcal{A}_{\alpha\beta}\left(  \tau\right)
\equiv~\left\langle \xi_{\alpha}\left(  \tau\right)  \xi_{\beta}\left(
0\right)  -\xi_{\beta}\left(  \tau\right)  \xi_{\alpha}\left(  0\right)
\right\rangle ^{\ast}$ with any $\tau$. Since most correlations decay with
time, we expect each term of $\mathcal{A}$ to vanish for large $\tau$. Yet,
there are situations where it may \textit{increase} with $\tau$ up to some
maximum at $\hat{\tau}$, before decaying\cite{ShrZ14}. Not only would
$\mathcal{A}_{\alpha\beta}\left(  \hat{\tau}\right)  $ provide a more
prominent signal of TRA than $\mathcal{L}$, it points to the presence of a
characteristic time scale. The implications of $\hat{\tau}$ for systems with
\textquotedblleft subtle\textquotedblright\ signals are clearly worthy of
pursuit.\ (ii) We noted that $Q\left(  \mathcal{L}\right)  $ undergoes a
transition from an asymmetric Laplacian-like distribution (with exponential
decays) to an approximate Gaussian (with zero mean) as the system makes a Hopf
bifurcation. How does the linear behavior in $\ln Q$ crossover to the
quadratic? via anomalous exponents or at some crossover value $\ell_{\times}$?
If the latter, how does it vary with critical control parameter(s)? For simplicity, 
we presented cases with just two variables. Would a \textquotedblleft
many-body\textquotedblright\ system provide us with further, unexpected
collective behaviour? (iii) We showed that $\mathcal{L}$ (and $\tilde{\Gamma}$
for more restrictive systems) serves as an effective measure for detecting TRA
in simple model systems. Beyond these immediate questions, does this line 
of inquiry lead to any new insights
for other NESS in complex systems in nature, such as those in chemical
reactions, biological sciences, socio-economic arenas, global
ecology/climate, and stellar interiors? Most crucially, our hope is that it
provides an inroad into formulating an overarching framework for statistical
mechanics of non-equilibrium steady states.

\newpage

\begin{acknowledgments}
It is a pleasure to dedicate this article to Uwe T\"{a}uber's 60th birthday,
as the event reminded the author of many productive hours of enlightening
discussions on non-equilibrium statistical mechanics with him. The author 
gratefully acknowledges numerous similar discussions and valuable insights on
this topic with his collaborators, most recently R. Dickman, M.O.
Lavrentovitch, and J.B. Weiss. He also thanks E.F. Redish for help preparing
the manuscript. 

\end{acknowledgments}

\newpage

\appendix

\section{Decomposition of $\left\langle \xi_{\alpha}v_{\beta}\right\rangle
^{\ast}$}
\label{ap:xv=D+L}

In the steady state, $\left\langle \xi_{\alpha}\xi_{\beta}\right\rangle
^{\ast}$ at one time is the same as at an infinitesimal time later (when each
$\xi$ changes by $\varepsilon\partial_{t}\xi=\varepsilon v$). Let us write
this statement as%
\begin{equation}
\left\langle \xi_{\alpha}\xi_{\beta}\right\rangle ^{\ast}=\left\langle \left(
\xi_{\alpha}+\varepsilon v_{\alpha}\right)  \left(  \xi_{\beta}+\varepsilon
v_{\beta}\right)  \right\rangle ^{\ast} \label{TTI}%
\end{equation}
Using the Langevin representation, we have%
\[
v_{\alpha}=\mu_{\alpha}+\eta_{\alpha}%
\]
where%
\[
\left\langle \eta_{\alpha}\right\rangle =0;~~\left\langle \eta_{\alpha}%
\eta_{\beta}\right\rangle =2D_{\alpha\beta}\frac{1}{\varepsilon}%
\]
with $1/\varepsilon$ representing $\delta$ in $\left\langle \eta_{\alpha}%
\eta_{\beta}\right\rangle \propto\delta\left(  t-t^{\prime}\right)  $. In the
the limit of $\varepsilon\rightarrow0$, we find from (\ref{TTI})%
\[
\left\langle \xi_{\alpha}v_{\beta}+v_{\alpha}\xi_{\beta}\right\rangle ^{\ast
}=-\varepsilon\left\langle \eta_{\alpha}\eta_{\beta}\right\rangle
=-2D_{\alpha\beta}%
\]
Here, we verify that the units of both sides are indeed $\left[  \xi_{\alpha
}\right]  \left[  \xi_{\beta}\right]  \left[  t\right]  ^{-1}$. Meanwhile, by
definition, we have%
\[
\left\langle \xi_{\alpha}v_{\beta}-v_{\alpha}\xi_{\beta}\right\rangle ^{\ast
}=\mathcal{L}_{\alpha\beta}%
\]
Defining%
\[
A_{\alpha\beta}\equiv\mathcal{L}_{\alpha\beta}/2
\]
we see that it is precisely the rate of \textquotedblleft
area\textquotedblright\ (in the $\alpha$-$\beta$ plane) being swept out. Then,
$\left\langle \xi_{\alpha}v_{\beta}\right\rangle ^{\ast}$ can be decomposed to
a symmetric part ($-D$) and an anti-symmetric one ($A$):%
\[
\left\langle \xi_{\alpha}v_{\beta}\right\rangle ^{\ast}=-D_{\alpha\beta
}+A_{\alpha\beta}%
\]
In this sense, we regard the probability angular momentum to be on the same
footing as the diffusion coefficients. Finally, since $\left\langle
\eta\right\rangle =0$, we can also write the decomposition as $\left\langle
\xi_{\alpha}\mu_{\beta}\right\rangle ^{\ast}=-D_{\alpha\beta}+A_{\alpha\beta}%
$. For LGMs, $\mu_{\beta}=-F_{\beta}^{\gamma}\xi_{\gamma}$, so that we end up
with $F_{\beta}^{\gamma}C_{\gamma\alpha}=D_{\alpha\beta}-A_{\alpha\beta}$,
where $C_{\gamma\alpha}=\left\langle \xi_{\alpha}\xi_{\gamma}\right\rangle
^{\ast}$. Since $D$ is symmetric, we can summarize neatly%
\begin{equation}
F_{\alpha}^{\gamma}C_{\gamma\beta}=D_{\alpha\beta}+A_{\alpha\beta}
\label{FC=D+A}%
\end{equation}

\section{A Simple 3-state system in NESS}

\label{ap:3st} Here we include some details of the system with $i=1,2,3$
configurations (\textquotedblleft micro-states\textquotedblright) presented in
subsection \ref{3st}. Writing the probabilities to find the system in $i$ at
discrete time $\tau$ as a column vector,%

\[
\left\vert P\right\rangle \equiv\left(
\begin{array}
[c]{c}%
P\left(  1,\tau\right)  \\
P\left(  2,\tau\right)  \\
P\left(  3,\tau\right)
\end{array}
\right)
\]
the Master equation reads%
\[
\Delta_{\tau}\left\vert P\right\rangle =\mathfrak{L}\left\vert P\right\rangle
\]
where%
\[
\mathfrak{L}=\left(
\begin{array}
[c]{ccc}%
-\varphi-\tilde{\varphi} & 1 & 1\\
\varphi & -1-\varphi & 1\\
\tilde{\varphi} & \varphi & -2
\end{array}
\right)
\]
Conservation of probability is $\sum_{i}P\left(  i,\tau\right)  =1$,
consistent with $\left\langle w_{0}\right\vert \equiv\left(
\begin{array}
[c]{ccc}%
1 & 1 & 1
\end{array}
\right)  $ being a \textit{left} eigenvector of $\mathfrak{L}$ with zero
eigenvalue: $\omega_{0}=0$. The associated \textit{right} eigenvector is the
stationary distribution:%

\[
\left\vert P^{\ast}\right\rangle \equiv\left\vert u_{0}\right\rangle =\frac
{1}{Z}\left(
\begin{array}
[c]{c}%
2+\varphi\\
2\varphi+\tilde{\varphi}\\
\varphi^{2}+\tilde{\varphi}\varphi+\tilde{\varphi}%
\end{array}
\right)
\]
where $Z=\left(  2+\varphi\right)  \left(  1+\varphi+\tilde{\varphi}\right)
$. The other two ($n=1,2$) left and right eigenvectors are%

\begin{align*}
\left\langle w_{n}\right\vert  &  =\left(
\begin{array}
[c]{ccc}%
\varphi+\tilde{\varphi} & -1 & -1
\end{array}
\right)  ,~~\left(
\begin{array}
[c]{ccc}%
\tilde{\varphi}-\varphi^{2} & 2\varphi-\varphi\tilde{\varphi}-\tilde{\varphi}
& \varphi+\tilde{\varphi}-2
\end{array}
\right)  \\
\left\vert u_{n}\right\rangle  &  =\left(
\begin{array}
[c]{c}%
\tilde{\varphi}-1\\
-\left(  1+\varphi\right)  \\
\varphi-\tilde{\varphi}%
\end{array}
\right)  ,~~\left(
\begin{array}
[c]{c}%
0\\
1\\
-1
\end{array}
\right)
\end{align*}
associated with the eigenvalues:
\[
\omega_{n}=-\left(  1+\varphi+\tilde{\varphi}\right)  ,~~-\left(
2+\varphi\right)
\]
As a result, for such a simple system, we can provide the full, time
dependent, solution to our system:%
\begin{align*}
\left\vert P\right\rangle  &  =\left(  \mathfrak{I}+\mathfrak{L}\right)
^{\tau/\varepsilon}\left\vert P_{ini}\right\rangle \\
&  =%
{\displaystyle\sum\limits_{n=0}^{2}}
\left(  1+\omega_{n}\right)  ^{t/\tau}\frac{\left\langle w_{n}\right\vert
\left.  P_{ini}\right\rangle }{\left\langle w_{n}\right\vert \left.
u_{n}\right\rangle }\left\vert u_{n}\right\rangle \\
&  =\left\vert P^{\ast}\right\rangle +%
{\displaystyle\sum\limits_{n=1}^{2}}
\left(  1+\omega_{n}\right)  ^{t/\tau}\frac{\left\langle w_{n}\right\vert
\left.  P_{ini}\right\rangle }{\left\langle w_{n}\right\vert \left.
u_{n}\right\rangle }\left\vert u_{n}\right\rangle
\end{align*}
where $\left\vert P_{ini}\right\rangle $ is the initial distribution and
$\mathfrak{I}$ is the identity matrix.

\section{Analysis of $\tilde{Q}$}
\label{ap:Q}

A few details of the derivation of the behavior of $Q\left(  \mathcal{L}%
\right)  $ are provided here.

From Eqns. (\ref{Q},\ref{Q-ft}), we find its Fourier transform, associated
with $\alpha\beta$ component of the probability angular momentum, is given by
\[
\tilde{Q}\left(  z\right)  =\int\left[  \exp iz\left(  \xi_{\alpha
}\partial_{t}\xi_{\beta}-\xi_{\beta}\partial_{t}\xi_{\alpha}\right)  \right]
P\left(  \vec{\eta}\right)  P^{\ast}\left(  \vec{\xi}\right)  d\vec{\eta}%
d\vec{\xi}%
\]
Since the integrand is Gaussian (in $\vec{\xi},\vec{\eta}$), with
\[
iz\left(  -\xi_{\alpha}F_{\beta}^{\gamma}\xi_{\gamma}+\xi_{\beta}F_{\alpha
}^{\gamma}\xi_{\gamma}\right)  +iz\left(  \xi_{\alpha}\eta_{\beta}-\xi_{\beta
}\eta_{\alpha}\right)  -\eta_{\mu}\left(  4D_{\mu\nu}/\varepsilon\right)
^{-1}\eta_{\nu}-\xi_{\mu}\left(  2C_{\mu\nu}\right)  ^{-1}\xi_{\nu}%
\]
in the exponent, the integrals can be performed. To integrate over $\vec{\eta
}$, we defined $\mathbb{X}_{\left(  \alpha\beta\right)  }$, an anti-symmetric
matrix with elements%
\[
X_{\left(  \alpha\beta\right)  }^{\mu\nu}=\delta_{\alpha}^{\mu}\delta_{\beta
}^{v}-\delta_{\alpha}^{\nu}\delta_{\beta}^{\mu}%
\]
The subscript $\left(  \alpha\beta\right)  $ should be regarded as a label
rather than indices. Then the part of the exponent linear in $\vec{\eta}%
\,\ $is just $\xi_{\mu}\left(  izX_{\left(  \alpha\beta\right)  }^{\mu\lambda
}\right)  \eta_{\lambda}$, so that the result is proportional to $\exp\left[
-\xi_{\mu}G^{\mu\nu}\xi_{\nu}/2\right]  $, where%
\[
G^{\mu\nu}=\left(  C_{\mu\nu}\right)  ^{-1}+2izX_{\left(  \alpha\beta\right)
}^{\mu\lambda}F_{\lambda}^{\nu}+z^{2}X_{\left(  \alpha\beta\right)  }%
^{\mu\lambda}\left(  2D_{\lambda\sigma}/\varepsilon\right)  X_{\left(
\alpha\beta\right)  }^{\nu\sigma}%
\]

Using the notation of $\mathbb{D}$,$\mathbb{F}$, etc., we find the $\vec{\xi}$
integration leading to $\tilde{Q}=\left[  \det\mathbb{C}^{-1}%
/\det\mathbb{G}\right]  ^{1/2}$ (with $\det\mathbb{C}^{-1}$ coming from the
normalization of $P^{\ast}$), i.e.,
\begin{equation}
\tilde{Q}\left(  z\right)  =\left\{  \det\left[  \mathbb{I}%
+2iz\mathbb{CXF}+\left(  2z^{2}/\varepsilon\right)  \mathbb{CXDX}^{T}\right]
\right\}  ^{-1/2} \label{QQQQ}%
\end{equation}
Here, the elements of $\mathbb{CXF}$ are%
\[
C_{\mu\sigma}X_{\left(  \alpha\beta\right)  }^{\sigma\lambda}F_{\lambda}^{\nu
}=C_{\mu\alpha}F_{\beta}^{\nu}-C_{\mu\beta}F_{\alpha}^{\nu}%
\]
Thus, its trace is
\[
Tr\mathbb{CXF}=F_{\beta}^{\gamma}C_{\gamma\alpha}-F_{\alpha}^{\gamma}%
C_{\gamma\beta}=-\left\langle \mathcal{L}_{\alpha\beta}\right\rangle ^{\ast}%
\]
so that the first two terms of the Taylor series for $\tilde{Q}$
($\tilde{Q}\left(  0\right)  =1$ and $\partial_{z}\tilde{Q}\left(  0\right)  =i\left\langle \mathcal{L}_{\alpha\beta}\right\rangle
^{\ast}$) are correct.

To obtain the full $Q\left(  \mathcal{L}\right)  $, we must perform the
inverse Fourier transform, $\int\tilde{Q}\left(  z\right)
e^{-iz\mathcal{L}}dz$, by exploiting the singularities of 
$\tilde{Q}\left(  z\right)  $. Since $\mathbb{X}$ is rank 2, the matrices in 
Eqn.(\ref{QQQQ}) are also effectively $2\times2$ matrices. Thus, we have exactly%
\[
\det\left(  \mathbb{I+M}\right)  =1+Tr\mathbb{M}+\det\mathbb{M}%
\]
and all we need are the two eigenvalues of $\mathbb{M}$. In our case, the
matrix elements are real and quadratic in $iz$ , so that $\left(  \tilde
{Q}^{\ast}\right)  ^{-2}$ is a quartic polynomial, \textit{real} in $iz$.
Thus, it is proportional to $\left(  iz-\chi_{+}\right)  \left(  iz-\chi
_{+}^{\ast}\right)  \left(  iz-\chi_{-}\right)  \left(  iz-\chi_{-}^{\ast
}\right)  $ and the singularities of $\tilde{Q}$ are branch points at
$iz=\chi_{\pm}$ and $\chi_{\pm}^{\ast}$. Typically, $\operatorname{Re}\chi
_{+}$ and $\operatorname{Re}\chi_{-}$ are of opposite signs and, iff DB is
obeyed, $\operatorname{Re}\chi_{+}=-\operatorname{Re}\chi_{-}$. (For
convenience, let us choose $\operatorname{Re}\chi_{+}>0$.) As a result, these
points lie in the four different quadrants of the complex $z$ plane. For $\int
dz$, we can choose two branch cuts joining the two c.c. pairs (in $iz$), i.e.,
from $z=-\operatorname{Im}\chi_{\alpha}-i\operatorname{Re}\chi_{\alpha}$ to
$z=+\operatorname{Im}\chi_{\alpha}-i\operatorname{Re}\chi_{\alpha}$
($\alpha=\pm$). As an example, $\chi_{+}=\left(  31.1+0.984i\right)
\times10^{-3}$ and $\chi_{-}=\left(  -32.1+1.02i\right)  \times10^{-3}$ in the
case associated with Fig.\ref{fig:LGM-focus}. By choosing the cuts this way,
the contour can be closed in either the upper or the lower half plane,
depending on the sign of $\mathcal{L}$. Deforming these contours to wrap
around the cut(s), we find that $\operatorname{Re}\chi_{\alpha}$ will control
the large $\mathcal{L}$ behavior of $Q$. If the signs of the
$\operatorname{Re}\chi_{\alpha}$'s are opposite, the distribution $Q\left(
\mathcal{L}\right)  $ will be dominated by exponential decays as
$\mathcal{L}\rightarrow\pm\infty$ with coefficient $\left\vert
\operatorname{Re}\chi_{_{\pm}}\right\vert $. (If the signs of the
$\operatorname{Re}\chi_{\alpha}$'s were the same, then $Q\left(
\mathcal{L}\right)  $ would vanish identically for $\mathcal{L}%
\operatorname{Re}\chi_{\alpha}<0$.) Typically, the rest of the contour
integral around the cut contributes to more slowly varying aspects of
$Q\left(  \mathcal{L}\right)  $. In the example quoted above, the dominant
parts of $\ln Q$ are predicted to be $-0.0311\mathcal{L}$ for $\mathcal{L}>0$
and $0.0321\mathcal{L}$ for $\mathcal{L}<0$. Comparing with the simulation
results shown in Fig.\ref{fig:LGM-focus}b,c, we find that the drops in $\ln Q$
are indeed almost linear (with slightly different slopes). All the
quantitative aspects are entirely consistent with the theoretical predictions.

\section{Linear Gaussian models with stable focus}
\label{ap:LGM-focus}

Here, we show that, in LGMs, DB is always violated if the deterministic force
takes the system to a stable focus. In the main text, we noted that DB is
satisfied iff $\left(  D^{-1}\right)  _{\alpha\beta}F_{\gamma}^{\beta}$ is
symmetric and positive definite. So, we just need show this product is not
symmetric if any eigenvalue of $F_{\gamma}^{\beta}$ is complex (with positive
real part). We are not aware if this question was answered for arbitrary $N$.
Here, we provide an explicit result for $N=2$ matrices, for which we exploit
the representation in terms of Pauli matrices, $\bbsigma$.

Since $\mathbb{D}$ is real symmetric and positive definite, let us write%
\[
\mathbb{D}=d_{0}\bbsigma_{0}+d_{1}\bbsigma_{1}+d_{3}\bbsigma_{3}%
\]
with real $d$'s and%
\[
d_{0}>\sqrt{d_{1}^{2}+d_{3}^{2}}%
\]
as well as%
\[
\mathbb{D}^{-1}=\frac{d_{0}\bbsigma_{0}-d_{1}\bbsigma_{1}-d_{3}\bbsigma_{3}%
}{\det\mathbb{D}}%
\]
Meanwhile, $\mathbb{F}$ is real but not symmetric in general, so that%
\[
\mathbb{F}=f_{0}\bbsigma_{0}+f_{1}\bbsigma_{1}+f_{3}\bbsigma_{3}%
+f_{2}\bbsigma_{2}%
\]
with real $f_{0,1,3}$ and $if_{2}$. To generate a stable focus, its
eigenvalues must be complex conjugate pairs with positive real parts, i.e.,%
\[
f_{0}>0;~~\left\vert f_{2}\right\vert ^{2}>f_{1}^{2}+f_{3}^{2}%
\]
Thus, the anti symmetric part of $\mathbb{D}^{-1}\mathbb{F}$ (the coefficient
of $i\sigma_{2}$) is
\begin{equation}
d_{0}\left[  if_{2}\right]  +\left[  d_{1}f_{3}-d_{3}f_{1}\right]  \label{asp}%
\end{equation}
But,%
\[
\left[  d_{1}f_{3}-d_{3}f_{1}\right]  ^{2}<\left(  d_{1}^{2}+d_{3}^{2}\right)
\left(  f_{1}^{2}+f_{3}^{2}\right)  <d_{0}^{2}\left\vert f_{2}\right\vert ^{2}%
\]
In other words, in expression (\ref{asp}), the magnitude of the first term is always
greater than that of the second. As a result, $\mathbb{D}^{-1}\mathbb{F}$ will
always have a non-vanishing anti-symmetric part, a signal of DB violation.

\section{LGMs from ENSO and MJO data}
\label{ap:clim}

Here, we provide the numerical values for the $\mathbb{D}$ and $\mathbb{F}$
matrices for the LGMs which best fitted to the time series of (a) NINO3-d20 in
El Ni\~{n}o and (b) the amplitudes of the principal components of two
empirical orthogonal functions of filtered outgoing long-wave radiation
associated with the Madden-Jullien Oscillations. For details of how these data
are collected and analyzed, as well as where they can be found, see Ref.
\cite{WFMNZ21}. For convenience, we quote only three significant figures here.

\begin{itemize}
\item El Ni\~{n}o:
\[
\mathbb{D}=\left(
\begin{array}
[c]{cc}%
0.0484\  & \ 8.09\\
8.09\  & \ 1.90\times10^{4}%
\end{array}
\right)  ,~~10^{3}\mathbb{F}=\left(
\begin{array}
[c]{cc}%
61.1\  & -0.194\ \\
125\  & 27.2\
\end{array}
\right)
\]
The units of NINO3 and d20 are $^{\circ}C$ and $cm$, while the time step
($\varepsilon$) is $month$. Thus, the units of $\left(  D_{11},D_{12}%
,D_{22}\right)  $ are $\left(  ^{\circ}C^{2},^{\circ}C\text{-}cm,cm^{2}%
\right)  month^{-1}$ , while all elements of $\mathbb{F}$ have unit
$month^{-1}$. Note that the numerical values of $\mathbb{D}$ appear to be
quite disparate. However, if we used $m$ instead of $cm$ for d20, then $8.09$
would be $0.0809$, while $D_{22}$ would be $1.90$. Obviously, we can choose 
units so that the diagonal elements of D appear as 1. Then the values of 
the off diagonal elements will provide a better sense of how strongly the 
two sets of noise are correlated. 

\item MJO:%
\[
10^{3}\mathbb{D}=\left(
\begin{array}
[c]{cc}%
5.97 & 0.225\\
0.225 & 6.71
\end{array}
\right)  ,~~10^{3}\mathbb{F}=\left(
\begin{array}
[c]{cc}%
8.32 & 102\\
-124 & 5.70
\end{array}
\right)
\]
The units of both variables here are the $amplitudes$ of the two principal
components while the time step ($\varepsilon$) is $day$. Thus, the units of
$\mathbb{D}$ and $\mathbb{F}$ are, respectively, $amplitude^{2}/day$ and
$day^{-1}$. Note that the eigenvalues of $\mathbb{F}$ have large imaginary
parts. Together with $\mathbb{D}$ being quite distinct from $\mathbb{I}$, this
system embodies both aspects of DB violation discussed in the main text.
\end{itemize}

\bigskip

\begin{thebibliography}{99}                                                                                               %


\bibitem {DZ18}Dickman R and Zia R\ K\ P 2018 Driven Widom-Rowlinson lattice
gas. \textit{Phys. Rev.} \textbf{E 97}, 062126

\bibitem {LDZ21}Lavrentovich M O, Dickman R and Zia R\ K\ P 2021
Microemulsions in the driven Widom-Rowlinson lattice gas. \textbf{E104}, 064135

\bibitem {Langevin}Langevin P 1908 Sur la th\'{e}orie du mouvement
brownien\textit{ C. R. Acad. Sci. Paris}. \textbf{146} 530; Kawasaki K 1973
Simple derivations of generalized linear and nonlinear Langevin equations
\textit{J. Phys. A: Math. Nucl. Gen.} \textbf{6} 1289. See also https://en.wikipedia.org/wiki/Langevin\_equation

\bibitem {Fokker}Fokker A D 1914 Die mittlere Energie rotierender elektrischer
Dipole im Strahlungsfeld\ \textit{Ann. Phys.} \textbf{348} 810; Planck M 1917
\"{U}ber einen Satz der statistischen Dynamik und seine Erweiterung in der
Quantentheorie \textit{Sitzungsber. Kgl. Preuss. Akad. Wiss.} \textbf{24} 324;
Risken H 1989 \textit{The Fokker-Planck equation: methods of solution and
applications} (Berlin: Springer). See also https://en.wikipedia.org/wiki/Fokker--Planck\_equation

\bibitem {PF-wiki}Perron O 1907 Zur Theorie der Matrices, \textit{Math. Ann.}
\textbf{64} 248; Frobenius G 1912 \"{U}ber Matrizen aus nicht negativen
Elementen, \textit{Sitzungsber. Kgl. Preuss. Akad. Wiss.} 456. See also https://en.wikipedia.org/wiki/Perron\%E2\%80\%93Frobenius\_theorem

\bibitem {W11}Wegscheider R 1911 \"{U}ber simultane Gleichgewichte und die
Beziehungen zwischen Thermodynamik und Reactionskinetik homogener Systeme
\textit{Monatshefte f\"{u}r Chemie}, \textbf{32} 849

\bibitem {Komo36}Kolmogoroff, A 1936 Zur Theorie der Markoffschen Ketten.
\textit{Math. Ann}. \textbf{112} 115

\bibitem {ZS07}Zia, R K P and Schmittmann B 2007 Probability currents as
principal characteristics in the statistical mechanics of nonequilibrium
steady states \textit{J. Stat. Mech.} \textbf{2007} P07012

\bibitem {Hill66}Hill T L 1966 Studies in irreversible thermodynamics iv.
diagrammatic representation of steady state fluxes for unimolecular systems.
\textit{J. Theor. Biol}. \textbf{10} 442

\bibitem {WFMNZ21}Weiss J B, Fox-Kemper B, Mandal D, Nelson A D and Zia R K P
2020 Nonequilibrium oscillations, probability angular momentum, and the
climate system \textit{J. Stat. Phys.} \textbf{179}, 1010

\bibitem {Penland93}Penland C and Magorian T 1993 Prediction of Nino 3 sea
surface temperatures using linear inverse modeling \textit{J. Clim.}
\textbf{6}(6), 1067

\bibitem {MMS}Mori F, Majumdar S N and Schehr G 2021 Detecting nonequilibrium
dynamics via extreme value statistics \textit{Europhys. Lett}. \textbf{135}, 30003

\bibitem {MLax60}Lax M 1960 Fluctuations from the nonequilibrium steady
state\textit{ Rev. Mod. Phys}. \textbf{32}(1), 25

\bibitem {UCT98}T\"{a}uber U C and Zia R K P 1998 (unpublished). See also
Dotsenko V, Macio\l ek A, Vasilyev O and Oshanin G 2013 Two-temperature
Langevin dynamics in a parabolic potential \textit{Phys. Rev.} \textbf{E87}, 062130

\bibitem {JBW03}Weiss J B 2003 Coordinate invariance in stochastic dynamical
systems \textit{Tellus} \textbf{A 55}(3), 208

\bibitem {SatyaReset}Evans M R, Majumdar S N and Schehr G 2020 Stochastic
resetting and applications \textit{J. Phys. A: Math. Theor. }\textbf{53} 193001

\bibitem {BBF16}Battle C., Broedersz C.P., Fakhri N., Geyer V.F., Howard J.,
Schmidt C.F., and MacKintosh F.C. 2016 Broken detailed balance at mesoscopic
scales in active biological systems \textit{Science} \textbf{352}, 604

\bibitem {Hopf43}Hopf E 1943 Abzweigung einer periodischen L\"{o}sung von
einer station\"{a}ren L\"{o}sung eines Differentialsystems \textit{Ber. Verh.
S\"{a}chs. Akad. Wiss. Leipzig, Math.-Naturw. Kl.} \textbf{94}, 3. \ See also
Jackson E.A. 1989 \textit{Perspectives of Nonlinear Dynamics}, Vol. 1,2
(Cambridge: Cambridge University Press) and http://www.scholarpedia.org/article/Andronov-Hopf\_bifurcation.

\bibitem {SZ95}Schmittmann B and Zia R K P 1995 \textit{Statistical Mechanics
of Driven Diffusive Systems}. Vol. 17 of \textit{Phase Transitions and
Critical Phenomena}, eds. Domb C and Lebowitz J L (London: Academic).

\bibitem {YL52}Yang C\ N and Lee T D 1952 Statistical theory of equations of
state and phase transitions. I. Theory of condensation \textit{Phys. Rev.}
\textbf{87}, 404; Lee T D and Yang C\ N 1952 Statistical theory of equations
of state and phase transitions. II. Lattice gas and Ising model \textit{Phys.
Rev.} \textbf{87}, 410

\bibitem {KLS84}Katz S, Lebowitz J L, and Spohn H 1983 Phase transitions in
stationary nonequilibrium states of model lattice systems \textit{Phys. Rev.}
\textbf{B28} 1655; Katz S, Lebowitz J L, and Spohn, H 1984 Nonequilibrium
steady states of stochastic lattice gas models of fast ionic
conductors\ \textit{J. Stat. Phys.} 34 497.

\bibitem {DS95}Dickman R and Stell G 1995 Critical behavior of the
Widom--Rowlinson lattice model. \textit{J. Chem. Phys.} \textbf{102}, 8674

\bibitem {WR70}Widom B and Rowlinson J S 1970 New model for the study of
liquid-vapor phase transitions. \textit{J. Chem. Phys.} \textbf{52}, 1670

\bibitem {ABC}Schmittmann B, Hwang K and Zia R K P 1992 Onset of spatial
structures in biased diffusion of two species \textit{Europhys. Lett.}
\textbf{19} 19; Bassler K E, Schmittmann B and Zia R K P 1993 Spatial
structures with nonzero winding number in biased diffusion of two species
\textit{Europhys. Lett.} \textbf{24} 115; Leung K-t and Zia R K P 1997
Drifting spatial structures in a system with oppositely driven species
\textit{Phys. Rev.} \textbf{E56}, 308; and Adams D\ A, Schmittmann B and Zia R
K P 2007 Coarsening of \textquotedblleft clouds\textquotedblright' and dynamic
scaling in a far-from-equilibrium model system \textit{Phys. Rev.}
\textbf{E75}, 041123

\bibitem {Conway}Gardner M 1970 The fantastic combinations of John Conway's
new solitaire game `life' Scientific American. 223 120. See also http://www.scholarpedia.org/article/Game\_of\_Life

\bibitem {ShrZ14}Shkarayev M S and Zia R K P 2014 Exact results for a simple
epidemic model on a directed network: explorations of a system in a
nonequilibrium steady state. \textit{Phys. Rev.} \textbf{E 90} 032107
\end{thebibliography}
\end{document}